# Ribosome profiling reveals post-transcriptional buffering of divergent gene expression in yeast

Joel McManus[1*], Gemma May[1], Pieter Spealman[1], and Alan Shteyman[1]

1) Carnegie Mellon University, Department of Biological Sciences, Pittsburgh, PA 15213

*Corresponding Author

**Abstract**

Understanding the patterns and causes of phenotypic divergence is a central goal in evolutionary biology. Much work has shown that mRNA abundance is highly variable between closely related species. However, the extent and mechanisms of post-transcriptional gene regulatory evolution are largely unknown. Here we used ribosome profiling to compare transcript abundance and translation efficiency in two closely related yeast species (*S. cerevisiae* and *S. paradoxus*). By comparing translation regulatory divergence to interspecies differences in mRNA sequence features, we show that differences in transcript leaders and codon bias substantially contribute to divergent translation. Globally, we find that translation regulatory divergence often buffers species-differences in mRNA abundance, such that ribosome occupancy is more conserved than transcript abundance. We used allele-specific ribosome profiling in interspecies hybrids to compare the relative contributions of *cis*- and *trans*-regulatory divergence to species differences in mRNA abundance and translation efficiency. The mode of gene regulatory divergence differs for these processes, as *trans*-regulatory changes play a greater role in divergent mRNA abundance than in divergent translation efficiency. Strikingly, most genes with aberrant transcript abundance in F1 hybrids (either over- or under-expressed compared to both parent species) did not exhibit aberrant ribosome occupancy. Our results show that interspecies differences in translation contribute substantially to the evolution of gene expression. Compensatory differences in transcript abundance and translation efficiency may increase the robustness of gene regulation.

## Introduction

The relationship between genotype and phenotype is a long-standing biological puzzle. Genetic mutations can affect phenotypes both by changing the coding sequence of proteins and by changing the regulation of gene expression. As much as forty years ago, it became clear that differences in protein coding sequence are too infrequent to explain differences in phenotype(King and Wilson 1975). Consequently, evolution of gene regulation has been suggested as a major source of phenotypic diversity (Rockman and Kruglyak 2006; Carroll 2005).

With the advent of DNA microarrays in the 1990s, many researchers began examining gene expression differences between species. mRNA abundance levels were found to vary greatly both among and between biological species, including yeast (Brem et al. 2002; Tirosh et al. 2011; Bullard et al. 2010), fruit flies (Ranz et al. 2003; Rifkin et al. 2003), mice(Schadt et al. 2003), and humans (Pickrell et al. 2010). In general, these studies report that polymorphic mRNA abundance is present in at least 25% of genes in a species and mRNA levels are even more divergent between closely related species. However, a few recent studies suggest that protein abundance is more conserved than mRNA abundance across all domains of life, including *E. coli*, *S. cerevisiae*, and humans (Schrimpf et al. 2009; Laurent et al. 2010). Because these studies depended upon orthologous genes identifiable across vast evolutionary distance, their results could be biased towards high abundant, housekeeping genes. Nonetheless, these works suggest that post-transcriptional processes, including mRNA translation and protein turnover, could act to reduce the effects of variation in mRNA abundance and offset divergent gene expression.

Gene expression requires numerous steps, including transcription, mRNA translation, and protein turnover. Each of these processes is regulated through complex networks of *cis*-regulatory sequence elements and *trans*-acting factors. *cis*-regulatory elements (e.g. transcription factor binding sites and miRNA target sites) are localized to the alleles they regulate and thus affect gene expression in an allele-specific manner. In contrast, *trans*-acting factors (e.g. transcription factors, RNA binding proteins, and miRNAs) can affect the expression of both alleles in a diploid cell. Consequently, the roles of these network components can be dissected by comparing allele-specific gene expression in F1 hybrids. In F1 hybrids, two parental alleles are subjected to identical *trans*-regulatory environments. Thus, differences in gene expression from the two alleles reflect divergent activity of their *cis*-regulatory sequences. *trans*-regulatory divergence can then be inferred by comparing interspecific expression differences with allele-specific expression differences in F1 hybrids (Wittkopp et al. 2004; Tirosh et al. 2009). This hybrid approach has revealed that most intraspecies variation in mRNA abundance results from *trans*-regulatory differences, but *cis*-regulatory differences accumulate over evolutionary time (Wittkopp et al. 2008a; Lemos et al. 2008; Emerson et al. 2010).

Divergence of gene regulatory networks has also been implicated in inheritance of gene expression levels. Like any phenotype, mRNA abundance can be inherited either additively or non-additively. One extreme form of non-additive inheritance includes cases in which genes are misexpressed in F1 hybrids, compared to parent species. This misexpression includes both over- and under-dominant inheritance, in which total mRNA abundance in F1 hybrids is either much higher or lower than both parent species, respectively. Misexpression of mRNA abundance can affect many genes, (Ranz et al. 2004; Gibson et al. 2004), and is associated with antagonistic *cis*- and *trans*-regulatory changes (Landry 2005; McManus et al. 2010; Schaefke et al. 2013).

While the evolution of mRNA abundance has been studied extensively, divergent regulation of mRNA translation has received much less attention. We used ribosome profiling to compare mRNA abundance and translation efficiency in two species of yeast, *S. cerevisiae* and *S. paradoxus*. Interspecies differences in mRNA abundance and translation efficiency affected similar proportions of the transcriptome. Importantly, differences in translation efficiency tend to buffer differences in mRNA abundance. We further compared the relative contributions of *cis*- and *trans*-regulatory differences in mRNA abundance and translation efficiency using F1 hybrids. Differences in both mRNA levels and translation efficiency are largely attributed to *trans*-regulatory divergence. Finally, we investigated inheritance patterns of mRNA abundance and ribosome occupancy in F1 Hybrids. Strikingly, ribosome occupancy was much less likely to exhibit misexpression, especially for genes involved in cell cycle regulation. These results reveal the importance of translation regulation in evolution, provide insights into the genetic and molecular mechanisms of translation regulatory divergence, and indicate that translation regulatory networks often act as a buffer to gene regulatory divergence in yeast.

## Results

### Measuring translation efficiency with ribosome profiling.

To investigate translation regulatory divergence, we used ribosome profiling in diploid *S. cerevisiae*, *S. paradoxus*, and their F1 hybrid (Fig 1A). This technique provides estimates of the relative number of ribosomes translating mRNA from each gene (ribosome occupancy). Compared with RNA-seq or microarray measurements of mRNA abundance, ribosome occupancy is a better proxy of protein abundance in yeast (Ingolia et al. 2009). This is consistent with the common model that yeast protein synthesis is largely regulated at the translation initiation step(Shah et al. 2013; Lackner et al. 2007; Sonenberg and Hinnebusch 2009). Ribosome bound mRNA were extracted from cycloheximide treated cells and digested with RNase I to produce short (28-30 nucleotides) Ribosome protected mRNA fragments (RPFs). RPFs were then purified and used to generate strand-specific libraries for Illumina high-throughput sequencing (Ingolia et al. 2011). Strand-specific polyA-positive mRNA-seq libraries were made in parallel. Together, this approach measures both ribosome occupancy (RPF) and mRNA abundance (mRNA) that can be compared to estimate relative translation efficiency, defined here as the ratio of ribosome footprints to mRNA fragments (RPF / mRNA)(Ingolia et al. 2009).

We developed an analysis pipeline to compare translation efficiency between species and between alleles in F1 hybrids that leverages the high quality genome sequences and gene annotations from each species (Scannell et al. 2011). To improve the annotations of introns and translation start sites, we sequenced the transcriptome of *S. paradoxus* (Scannell et al. 2011). Our reannotation of the *S. paradoxus* genome resulted in a working set of 5,474 orthologous protein coding genes. Sequence reads from each species were aligned to their respective genomes. Reads from F1 hybrid samples were aligned to each species genome separately, and alignments were compared to identify reads that map preferentially to one allele (McManus et al. 2010) (Fig 1B). This analysis resulted in an average of 7.4 million RPF and 6.9 million mRNA reads covering ORFs in each replicate (Table S1).

In some cases, allele-specific alignment can be biased towards higher quality genomes (Degner et al. 2009). However, these effects appear to be largely absent when using two high-quality genomes, as was the design of this study (McManus et al. 2010; Stevenson et al. 2013; Graze et al. 2012). To investigate the effectiveness of our allele-specific mapping pipeline, we combined sequence data from individual species to create a “mock hybrid”, and compared divergence estimates from allele-specific and single-species alignment procedures. Estimates of gene expression differences from allele-specific and single-species alignment procedures were highly consistent for ribosome occupancy (RPFs; Pearson’s $R^2 > 0.97$), mRNA abundance ($R^2 > 0.98$) and translation efficiency (RPF / mRNA; $R^2 > 0.96$) (Fig 1C and Fig S1), suggesting that the allele-specific analysis pipeline generates accurate estimates of regulatory divergence. We therefore used our allele-specific pipeline for both interspecies and F1 hybrid allelic comparisons.

**Interspecies differences in translation efficiency and mRNA abundance.**

We compared interspecies differences in mRNA abundance and ribosome occupancy using a custom analysis pipeline (see methods). Count-based tests were applied to identify statistically significant differences using 5% false discovery rate (FDR) and 1.5-fold minimum thresholds in both replicates. Our biological replicates were well correlated for both mRNA and RPF libraries ($\rho > 0.97$; Fig S2). By requiring at least 20 total allele-specific mapped reads (*S. cer* reads + *S. par* reads > 20) for both RPF and mRNA in each replicate, we identified 1,308 genes (26.9%) with differences in ribosome occupancy (RPFs), 1,739 genes (35.8%) with differences in mRNA abundance, and 1,345 genes (27.7%) with differences in translational efficiency (RPF/ mRNA) (Fig 2A-C). Similar results were obtained using minimum cutoffs of 10, 50, and 100 reads (Table S2). These results suggest that translation regulation may play as significant a role as mRNA abundance in the evolution of gene expression.

Many features of mRNA transcripts influence translation efficiency. Long transcript leaders (5' UTRs), frequent upstream AUGs (uAUGs), and uORF activity have all been associated with reduced translation efficiency (Rojas-Duran and Gilbert 2012; Waern and Snyder 2013; Brar et al. 2011; Ingolia et al. 2009). We used our RNA-seq data to annotate transcript leaders for 3,014 genes in each species. Species differences in transcript leader length were negatively correlated with divergent translation efficiency, such that the species with the longer transcript leader tended to have lower efficiency of translation ($\rho = -0.27$; $P = 1.9 \times 10^{-13}$). Differences in the number of uAUGs were also negatively correlated with divergent translation efficiency, suggesting that orthologs with more uAUGs tend to be less efficiently translated (ρ = -0.27; P = 6.3x10^-8). Partial correlation analysis of transcript leader length, controlling for uAUG frequency, and vice versa, show that they contribute equally to divergent translation efficiency (ρ = -0.126; P=0.00054; and ρ = -0.129; P = 0.00038). In comparison, the frequency of upstream stop codons was not correlated with divergent translation efficiency (P > 0.3). Thus, divergent translation regulation is due in part to divergent transcript leader length and uAUG frequency.

The importance of uAUGs suggests that upstream open reading frames (uORFs) may also contribute to divergent translation efficiency. Once thought to be rare, recent studies have identified thousands of uORFs in *S. cerevisiae*, many of which appear to initiate translation at non-canonical "NTG" start codons (Brar et al. 2011; Ingolia et al. 2009). We searched for putative uORFs starting with NTG and ending with an in-frame stop codon in transcript leader regions from each species. RPF reads were mapped to uORF regions and the ratios of uORF to main ORF reads from each species were compared. Species differences in ribosome occupancy over uORFs were also negatively correlated with divergent translation efficiency (ρ = -0.2477153; P = 3.6 x10^-8), such that orthologs with higher uORF occupancy showed generally lower efficiency of main ORF translation (Fig S3). These results indicate that translation regulatory divergence can be attributed in part to interspecies differences in uORF presence and utilization.

Due to redundancy in the genetic code, codon usage bias can also contribute to regulation of translation efficiency(Plotkin and Kudla 2010). A previous comparison of nine yeast species revealed divergent codon usage bias for hundreds of orthologous ORFs (Man and Pilpel 2007). We calculated codon bias using the tRNA Adaptation Index (tAI) (Reis et al. 2004) to compare species differences in codon bias to translation regulatory differences. Codon bias was positively correlated to mRNA abundance, ribosome occupancy, and translation efficiency( $\rho > 0.52$; $P < 2*10^{-16}$), such that highly expressed genes are biased towards using more abundant tRNA in both species. In contrast, interspecies differences in codon bias were weakly correlated with divergent translation efficiency ($\rho = 0.11$, $P = 4.706e-15$).

**Translation regulatory divergence buffers species differences in mRNA abundance**

Genes with interspecies differences in mRNA abundance were more than twice as likely to have divergent translation efficiency, compared to genes with conserved mRNA abundance (Fisher's Exact Test (FET), $P < 10^{-16}$). The co-occurrence of divergence in translation efficiency could work in the same (amplifying) or opposite (buffering) direction (Fig 3A). Strikingly, we found that buffering effects were ~5.5 times more common (N=560) than amplifying effects (N=101; FET, $P<10^{-16}$). Indeed there was a significant negative correlation between mRNA abundance and translation efficiency genome-wide (Fig 3B-C; $\rho = -0.413$, $P < 10^{-16}$). This tendency towards buffering would reduce divergent expression overall at the protein level.

We investigated several features of buffered and amplified genes. Highly expressed genes often have housekeeping functions, and therefore might be buffered more frequently than genes with low mRNA abundance. However the mRNA abundance of buffered genes was either average (in *S. cerevisiae*) or slightly lower than average (*S. paradoxus*) (Fig S4). We also compared the biological functions of buffered and amplified genes. Gene ontology (GO) analysis (Table S3) revealed that buffered genes were enriched in processes involved in cellular communication and catabolism ($P < 0.01$), while amplified genes, were enriched in functions related to more specific metabolic processes, including amino-acid metabolism, sulfur assimilation ($P < 0.01$), iron import, and redox reactions ($P < 0.004$). The enrichment of these categories is consistent with species / strain differences in metabolic strategies, as *S. paradoxus* metabolism uses aerobic respiration more intensely, while *S. cerevisiae* has more auxotrophies to amino-acid biosynthesis.

***cis-* and *trans*-contributions to translation regulatory divergence.**

To gain further insight into the molecular and genetic changes underlying translation regulatory divergence, we performed allele-specific ribosome profiling on an F1 hybrid strain made by crossing our strains of *S. cerevisiae* and *S. paradoxus*. In F1 hybrids, differences in allele-specific translation efficiency reflect divergence of their *cis*-regulatory sequences. The effects of *trans*-regulatory divergence can then be inferred by comparing inter-species expression differences with F1 hybrid allele-specific expression (Wittkopp et al. 2004; Tirosh et al. 2009; McManus et al. 2010; Bullard et al. 2010). This approach is feasible with our strains, as *S. cerevisiae* and *S. paradoxus* are sufficiently divergent to allow proper allele-assignment for 71% of sequence reads(Table S1). Figure 1C shows an example comparison of sequence coverage with separate species and allele-specific read assignments.

At each level of gene regulatory control - transcript abundance, ribosome occupancy, and translational efficiency - more genes were affected by differences in the activity of *trans*-regulatory factors than by *cis*-regulatory differences (Fig 4A). This is consistent with previous studies of regulatory divergence and shows that the mode of regulatory evolution is similar for mRNA abundance and translation efficiency. However, *cis*-regulatory divergence contributes more to differences in translation efficiency (36.7% *cis*) than to differences in mRNA abundance (30.2% *cis*) (Fig 4B; Wilcoxon Rank-sum Test (WRT); P = 3.9 x 10^-14).

We investigated the overlap of genes with *cis*-regulatory divergence in mRNA abundance, translation efficiency, and ribosome occupancy (Fig 4C). *cis*-regulatory divergence in translation efficiency was found three times more often in genes with *cis*-regulatory divergence in mRNA levels than compared to those without (FET, P < 2.2 x 10^-16). Coupled *cis*-regulatory evolution of mRNA abundance and translation efficiency was biased more than 3-fold towards buffering allele-specific ribosome occupancy. As a result, 345 genes (55%) with *cis*-regulatory differences in mRNA levels do not exhibit cis-regulatory divergence in ribosome occupancy. We refer to these genes as "*cis*-mRNA buffered", while genes with concurrent *cis*-regulatory differences in mRNA abundance and ribosome occupancy are "*cis*-mRNA penetrant". The median magnitude of cis-regulatory divergence for cis-mRNA buffered genes is significantly lower than that of cis-mRNA penetrant genes (1.84-fold vs 2.31-fold; WRT, P < 2.2 x 10^-16). This suggests that small magnitude *cis*-regulatory differences in mRNA levels are less likely to penetrate into the phenotype of protein production.

We next examined how *cis*- and *trans*-regulatory divergence in mRNA abundance and translation efficiency contribute to buffered and amplified interspecies expression differences. Compared to buffered genes, amplified genes were more likely to exhibit *cis*-regulatory divergence of both mRNA abundance (FET; 1.6 fold, P = 0.018) and translation efficiency (FET; 2.3-fold, P = 5.3 e-05). Amplified genes also had higher *%cis* than buffered genes, both for mRNA abundance (31.1% vs 25.5%; WRT P = 0.07) and translation efficiency (60.7% vs 31.7%; WRT P = 0.0003342; Figure 4). Thus, *cis*-regulatory divergence contributes more to amplified genes than to buffered genes.

**Buffering of misexpression in F1 hybrids**

Previous studies have revealed aberrant mRNA abundance in interspecies hybrids - either lower (underexpressed) or higher (overexpressed) than both parental species. This misexpression may be related to interspecies network incompatibilities and, potentially, to speciation (Ranz et al. 2004; Landry 2005; Moehring et al. 2007). However, a recent proteomics study estimated only 3% of 398 analyzed genes exhibit misexpression of protein abundance in F1 hybrids of *S. cerevisiae* and *S. bayanus* (Khan et al. 2012). We analyzed the inheritance of divergent expression in both mRNA abundance and ribosome occupancy to test the hypothesis that mRNA misexpression could be translationally buffered. Compared to mRNA abundance, ribosome occupancy showed strikingly fewer genes with non-additive inheritance modes (Figure 5), with 6.3-fold fewer underexpressed genes (FET P<10^-16) and 1.9-fold fewer overexpressed genes (FET; P= 0.0003194). This result indicates that divergent translation regulation also buffers misexpression of mRNA abundance, such that total protein synthesis in F1 hybrids is more consistent with that of parent species.

Buffering of misexpression could reduce hybrid incompatibilities. To investigate the effects of misexpression and buffering, we compared enrichment of GO functions in genes misexpressed at the mRNA level and at the level of ribosome occupancy. Genes underexpressed at the mRNA level were functionally enriched in cell cycle regulation and cellular metabolic processes, while genes with overexpressed mRNA largely functioned in purine biosynthesis pathways ($P < 0.01$; Table S4). Consistent with this, genes with aberrantly high ribosome occupancy in the F1 hybrid strain were enriched for functions in inosine monophosphate biosynthesis, the precursor to purine synthesis. In contrast, no GO functional categories were enriched in the 34 genes underexpressed at the translation level. In conclusion, compensatory changes in translation efficiency appear to correct aberrant mRNA abundance in F1 hybrids of *S. paradoxus* and *S. cerevisiae*, possibly restoring the expression of cell cycle control genes.

## Discussion

### The evolution of translation efficiency

Although undoubtedly important, evolutionary changes in mRNA abundance may not always reflect interspecies differences in gene expression. Indeed, prior work has highlighted a disconnect between variation in mRNA and protein abundance within yeast (Foss et al. 2007), mice (Ghazalpour et al. 2011), and humans (Wu et al. 2013). Together, these studies suggest that variation in mRNA translation and protein turnover contribute to polymorphic gene expression. Our results indicate that translation regulation plays a substantial role in the evolution of gene expression. Roughly one quarter of all expressed genes exhibited divergent translation efficiency, showing that divergence in translation regulation and mRNA abundance affect similar numbers of genes.

Changes in codon usage have been previously implicated in yeast gene regulatory evolution (Man and Pilpel 2007). More recent studies have highlighted important translation regulatory roles of transcript leaders in *S. cerevisiae* (Rojas-Duran and Gilbert 2012; Waern and Snyder 2013; Brar et al. 2011; Ingolia et al. 2009). Our results indicate that interspecies differences in transcript leaders appear to play a larger role than codon bias. One explanation for this observation is that changes in codon bias are more likely to be pleiotropic, as changes in codon usage can alter translation elongation and disrupt co-translational protein folding (Plotkin and Kudla 2010; Angov 2011). In contrast, changes in transcript leader length and sequence composition are likely to affect only translation initiation. Much like transcription promoter regions, transcript leaders provide important opportunities for evolutionary changes in gene expression without affecting protein-coding sequences.

**Differences in modes of regulatory evolution for mRNA abundance and translation efficiency.**

We used allele-specific ribosome profiling to compare the contributions of *cis*- and *trans*-regulatory changes toward divergence of mRNA abundance and translation efficiency. We found that *cis*-regulatory differences appear to contribute more to divergence in translation efficiency than to interspecies differences in mRNA abundance. This is inconsistent with other work suggesting fewer *cis*-acting loci contribute to polymorphic regulation of mRNA abundance than of protein abundance among *S. cerevisiae* strains (Foss et al. 2011). This may be due in part to a smaller sample size (354 genes) and lower power of mass spectrometry based studies for QTL mapping. In fact, recent work from the same group has shown that local pQTL are more common than previously appreciated (Albert et al., submitted). Differences in divergence time in the Foss study (between strains) and ours (between species) could also contribute to their differing conclusions, as it has been shown that *cis*-regulatory divergence accumulates over evolutionary time (Wittkopp et al. 2008b; Lemos et al. 2008; Emerson et al. 2010).

**Buffering of divergent mRNA abundance and translation efficiency**

Our results indicate that regulatory interactions between mRNA abundance and translation regulatory divergence more often reduce than amplify interspecies differences in gene expression. Buffering of gene expression may help explain recent work showing that protein abundance is more conserved than mRNA abundance across organisms (Schrimpf et al. 2009; Laurent et al. 2010). Changes in translation efficiency may compensate for changes in mRNA abundance by balancing the numbers of translating ribosomes.

There are two potential, non-exclusive mechanisms for buffering divergent gene regulation. First, regulatory networks controlling mRNA abundance and translation efficiency could have diverged through compensatory genetic mutations ("genetic compensation"), such that mutations in one pathway counteract the effects of mutations in a different pathway. Secondly, the gene regulatory networks may be inherently robust, such that changes in mRNA abundance are automatically buffered by feedback loops and regulatory circuitry("network robustness"). The contributions of *cis*- and *trans*-regulatory changes in buffered genes may help differentiate between these mechanisms of buffering. For example, the genetic compensation model predicts that buffered genes would more frequently exhibit counteracting *cis*-regulatory changes in mRNA abundance and translation efficiency. In contrast, counteracting *trans*-acting differences are more consistent with network robustness. Out of the 560 buffered genes in our dataset, 75 (13%) exhibit counteracting *cis*-regulatory changes while 254 (45%) have counteracting *trans*-acting changes. Consequently, we propose that network robustness is likely to be more common than genetic compensation.

Compared to buffered gene expression levels, amplified gene expression levels have fewer concordant *trans*-regulatory changes in both mRNA levels and translation efficiency. Of the 101 amplified genes, 23 (23%) have amplifying *cis*-regulatory changes in both mRNA abundance and translation efficiency, while 24 (24%) have amplifying *trans*-regulatory changes. Amplifying genes also had higher %*cis*, showing that *cis*-regulatory changes contribute more to divergent regulation of amplified genes, compared to buffered genes. Importantly, *cis*-regulatory divergence is thought to be more frequently mediated by positive selection than *trans*-regulatory divergence (Emerson et al. 2010; Graze et al. 2012; Schaefke et al. 2013). Thus amplifying divergence of mRNA abundance and translation efficiency may reflect positive selection for gene expression, as has been suggested for combinations of *cis*-regulatory mutations affecting mRNA abundance alone in *Drosophila* (Graze et al. 2012).

Many biological systems, including gene regulatory networks, are known to exhibit robustness to genetic and environmental perturbation (Levy and Siegal 2008; MacNeil and Walhout 2011; Masel and Siegal 2009; Denby et al. 2012; Stark et al. 2005). By increasing the robustness of protein production, translational buffering could reduce the phenotypic impacts of variation in mRNA abundance. This in turn would allow increased variation in mRNA abundance which, if unmasked by disruption of buffering, could give organisms more variation upon which selection might act. This would create a situation identical to canalization in development (Waddington 1942; Rutherford and Lindquist 1998), in which capacitor genes allow the buildup of latent variation which is only observed under extraordinary conditions. Our results suggest that translational buffering serves a similar purpose by allowing variation in mRNA abundance to accumulate through genetic drift without substantially changing protein levels.

Earlier studies on gene regulatory evolution have documented widespread misexpression of mRNA abundance in interspecies hybrids(Ranz et al. 2004; Gibson et al. 2004), often associated with counteracting *cis*- and *trans*-regulatory divergence of mRNA abundance (Landry 2005; McManus et al. 2010; Schaefke et al. 2013). Surprisingly, our analysis revealed translational buffering of misexpressed genes. Hundreds of genes had aberrant mRNA levels in hybrid cells, yet few showed aberrant ribosome occupancy patterns. As with interspecies differences in mRNA abundance,

over- and under-expressed genes could be buffered through genetic compensation or network robustness. It has been shown that protein-protein interactions are functional in F1 hybrids of *S. cerevisiae* and *S. kudriavzevii* (Leducq et al. 2012), suggesting that many biochemical features of regulatory networks are robust in interspecies hybrids. Thus, network robustness is an attractive answer. By reducing aberrant gene expression patterns, translational buffering of misexpression could contribute to the considerable success of yeast hybrids (Morales and Dujon 2012).

Thus far, we have investigated translation regulatory evolution only in the context of optimal growth conditions. Because many stress responses involve changes in mRNA translation, repeating these experiments under different conditions could shed light on the role of translation in species-specific responses to stress. It will also be important to identify the genes required to maintain translational buffering, as they could act as capacitors for variation in mRNA abundance. Although there is much yet to learn, our results underscore the importance of translation in the evolution of gene expression, help define the molecular and genetic causes of divergence in translation efficiency, and show that interspecies differences in translation efficiency often act as a buffer to gene regulatory evolution.

## Methods

### Yeast Strains

Saccharomyces strains: *S. cerevisiae* (GSY83; S288C), *S. paradoxus* (GSY82; CBS432), and their F1 hybrid (GSY88) were kind gifts from Gavin Sherlock (Kao et al. 2010). The original GSY83 is a haploid strain, while GSY82 and GSY88 are diploid strains. For this study, a diploid version of GSY83 was produced by galactose induced transient expression of the HO gene from plasmid YCP50-Gal::HO, a gift from John Woolford. Diploid GSY83 were identified via PCR screening.

**Ribosome profiling library preparation.**

Yeast polysome extracts were prepared for ribosome profiling as previously described (Ingolia 2010) with slight modifications (see supplemental methods). RNA for matched RNA-seq was extracted from 50µl of cell lysate using TRIzol (Life Technologies) and vortexing with glass beads. The sample was then enriched for mRNA using magnetic oligo-DT DynaBeads (Life Technologies) according to the manufacturers instructions. The resulting RNA was then fragmented by incubating in alkaline fragmentation buffer (50 mM $NaCO_3$, pH 9.2; 1 mM EDTA) at 95°C for 40 minutes, precipitated with one volume of isopropanol and resuspended in 20µl of nuclease free water.

High-throughput sequencing library preparation was performed as previously described (Ingolia et al. 2011) with slight modifications (see supplemental methods). RPF and mRNA library cDNA templates were amplified by 12 cycles of PCR using Phusion-polymerase™ (New England Biolabs), with primers incorporating barcoded Illumina TruSeq library sequences. The resulting PCR products were then purified using PCR purification columns (Qiagen) according to the manufacturers instructions. The quality and size of the PCR products were assessed using an Agilent Tapestation™. Libraries were sequenced on an Illumina HiSeq 2000 at the University of Southern California Epigenome Center.

**Sequence data alignment**

Raw sequences were trimmed of 3' adapters using custom perl scripts. For RPF data, reads with lengths of 27 to 33 nucleotides were retained for further analysis as this size is most likely to represent footprinted fragments. For mRNA, reads with lengths of 27 to 40 nucleotides were retained. Adapter trimmed reads were first aligned to genomes using Bowtie (Langmead et al. 2009) (version 0.12.8) with parameters --best --chunkmbs 500. Reads that failed to map to genomic regions were recovered and aligned to splice-junction databases requiring a minimum overlap of 6 nucleotides. Splice-junction read alignments were converted to SAM formatted genomic coordinates using a custom perl script.

Allele-specific read assignment was performed as described previously (McManus et al. 2010) with slight modifications. Briefly, adapter trimmed reads from separate species were concatenated into single "mixed parent" files. Mixed parent and hybrid sequence files were aligned to both species genome and junction databases as described above, and alignments were compared with a custom perl script to identify reads that map preferentially to one genome (Fig 1B). The library preparation procedure used in this frequently adds untemplated nucleotides during reverse transcription (Brar et al. 2012). Considering this, we ignored mismatches at the first two bases of the 5' end of sequence reads when comparing alignments between two species. Reads were assigned to the allele to which they map with the fewest mismatches. Estimates of interspecies differences in gene expression made through our allele-specific pipeline were highly correlated with those made by single-species alignment used in traditional

RNA-seq and ribosome profiling (Fig S1; $\rho > 0.96$ for RPF, mRNA and translation efficiency).

**Gene expression analyses**

Allele-specific reads were mapped to ORFs and uORFs and normalized by downsampling all genes from the species or allele with more mapped reads (McManus et al. 2010). We also accounted for species differences in ORF length by downsampling reads from the longer ortholog in proportion to the ratio of ortholog lengths (i.e. if one species' ORF was twice as long, we would divide the number of assigned reads by two). Tests of statistical significance were performed essentially as previously described (McManus et al. 2010) using R (version 2.15.2). For analyses of expression differences, raw p-values were corrected using the Benjamini-Hochberg method (Benjamini and Hochberg 1995), and significant differences were identified using 5% FDR and 1.5-fold magnitude thresholds. Significant differences in mRNA abundance and ribosome occupancy (RPF) were identified using binomial exact tests (BET), requiring both replicates meet significance thresholds. Differences in translation efficiency (RPF/mRNA) were identified using Cochran Mantel-Haenszel tests (CMH), which directly incorporate experimental replicates. CMH tests were used to evaluate *trans*-regulatory divergence by comparing the ratio of expression differences in parents and hybrids for mRNA abundance and ribosome occupancy. The method of Altman and Bland (Altman and Bland 2003) was used to test for *trans*-regulatory divergence in translation efficiency. Table S5 contains all read-count data, tAI calculations, expression ratios and results of significance test for divergent expression. For full expression analysis details, see the supplemental methods.

Genes were classified as “buffered” or “amplified” using the following criteria. Both buffered and amplified genes have significant differences in mRNA abundance and translational efficiency. In buffered genes, the interspecies difference in ribosome occupancy (RPF) is at least 1.5-fold lower than the interspecies difference in mRNA abundance. In contrast, amplified genes have at least 1.5-fold higher differences in ribosome occupancy compared to differences in mRNA abundance.

Categories of gene expression inheritance were determined using a custom R program as previously described (McManus et al. 2010). Total expression in F1 hybrid cells was normalized to that of parental samples and log-transformed. Genes with total expression more than 1.25-fold different from that of either parent species were considered to have nonconserved inheritance and were classified as having additive, dominant, under-dominant or over-dominant inheritance modes based on the magnitude of differences between hybrid and parental expression.

**Additional analyses**

Gene ontology enrichment analysis was performed using the generic gene ontology term finder (go.princeton.edu), and GO terms were further evaluated using revigo (revigo.irb.hr) (Supek et al. 2011). The tRNA adaptation index (tAI) was calculated using the codonR package (http://people.cryst.bbk.ac.uk/~fdosr01/tAI/index.html) (Reis et al. 2004). Genome browser tracks (bedGraph) were produced for visual analysis using genomeCoverageBed from bedTools. Images for all figures were produced using the Integrative Genome Browser (Robinson et al. 2011).

**Data access**

High-throughput sequencing data have been submitted to NCBI Sequence Read Archive (http://www.ncbi.nlm.nih.gov/sra) under accession number SRP028552. A read-count table for mRNA and RPFs, genome-browser tracks (bedGraph) and gene annotations (bed) have been submitted to NCBI Gene expression omnibus (http://www.ncbi.nlm.nih.gov/geo/) under accession number GSE52119.

**Acknowledgements**

We are grateful to Gavin Sherlock for sharing strains, and John Woolford and Adam Lindstedt for use of equipment. We also thank Zia Khan, Christine Vogel, and members of the McManus lab for comments and suggestions on the manuscript. This work was supported by laboratory start-up funding provided by the Department of Biological Sciences at Carnegie Mellon University.

**Author contributions**

CJM conceived of the experiments and designed the study. CJM, GEM, and PS performed the experiments. CJM and AS analyzed the data. CJM wrote the manuscript with input from all authors.

**Disclosure Declaration**

The authors are not aware of any conflicts of interest.

**Figure Legends**

**Figure 1.** Overview of allele-specific ribosome profiling to measure divergence in ribosome occupancy, mRNA abundance, and translational efficiency. (A) Ribosome profiling was performed on log-phase cultures of *S. cerevisiae*, *S. paradoxus*, and their F1 hybrid. Ribosome protected fragments (RPFs) were purified and cloned into Illumina high-throughput sequencing libraries (left). Poly-adenylated mRNA sequencing libraries were prepared in parallel (right). (B) Sequence reads from each sample were aligned to both species genomes. Allele-specific reads were identified by comparing genomic alignments and mapped to corresponding regions of orthologous ORFs. (C) Comparison of separate and allele-specific coverage in ribosome profiling experiments. IGV browser tracks showing normalized coverage of RPF and mRNA sequence reads from *S. cerevisiae* (blue) and *S. paradoxus* (magenta) over the *COX6* gene (YHR051W). Measurements of interspecies differences in read coverage are equivalent for separate species (upper) and allele-specific alignments (“mock hybrid”, lower; also see Fig S1).

**Figure 2.** Comparing regulatory divergence of ribosome occupancy, mRNA abundance, and translation efficiency. (A-C) Scatter plots compare the normalized average number of sequence reads for *S. cerevisiae* (x-axis) and *S. paradoxus* (y axis). Genes with statistically significant differences in read counts (FDR < 5%, minimum 1.5-fold difference) are plotted as open circles with black edges. Translation efficiency is defined here as the number of ribosome protected fragment reads (RPF) divided by the number of mRNA-seq reads covering an ORF.

**Figure 3.** Translation regulatory divergence buffers interspecies differences in mRNA abundance. (A) Cartoon depicting buffering (left) and amplification (right). mRNA are shown as blue lines, and ribosomes are shown as black circles. Buffered genes have divergent mRNA abundance, with less divergent ribosome occupancy such that protein production is more conserved. In contrast, amplified genes have divergent mRNA abundance and even more divergent ribosome occupancy. (B) Scatterplot comparing divergent translation efficiency (y-axis, RPF/mRNA) with divergent mRNA abundance (x-axis). Buffered and amplified genes are plotted in red and blue, respectively. The negative correlation between mRNA abundance and translation efficiency suggests a genome-wide trend towards buffering. (C) Example of RPF and RNA-seq coverage over *FPK1*, a buffered gene. IGV browser tracks showing normalized coverage for *S. cerevisiae* (blue) and *S. paradoxus* (red).

**Figure 4.** Contributions of *cis*- and *trans*-regulatory divergence in mRNA abundance, ribosome occupancy, and translation efficiency. (A) Bar plot shows the number of genes affected by significant regulatory divergence in *cis*-acting sequences (”C”) and *trans*-acting factors (”T”). *trans*-acting factors contribute most to mRNA abundance. (B) Box plot showing the fraction of regulatory divergence attributable to differences in *cis*-regulatory elements (%*cis*). Ribosome occupancy and translation efficiency both have higher %*cis* than mRNA abundance. (C) Venn diagram showing the overlap of genes with *cis*-regulatory divergence in mRNA abundance, ribosome occupancy, and translation efficiency. (D) Box plot depicting %*cis* for genes with buffering (”Buf”) and amplifying (”Amp”) regulatory divergence of mRNA abundance (left) and translation

efficiency (right). Asterisks indicate results of wilcoxon rank-sum test comparisons (*, $P < 0.001$; **, $P < 0.0005$).

**Figure 5.** Inheritance of gene expression in F1 hybrids of *S. cerevisiae* and *S. paradoxus*. (A) Hypothetical patterns of gene expression in *S. paradoxus* (blue), *S. cerevisiae* (red) and F1 hybrid yeast (purple) depicting six modes of gene expression inheritance. Scatterplots showing the difference between mRNA expression levels (left) in the F1 Hybrid and *S. cerevisiae* (x axis) and *S. paradoxus* (y axis). The difference between ribosome occupancy levels (RPF) in the F1 Hybrid and parental species is shown on the right. Bar plots shows the number of genes in each inheritance category from ribosome occupancy data. Genes with overdominant (overexpressed) and underdominant (underexpressed) inheritance occur much less frequently when considering ribosome occupancy, as compared to misexpression in mRNA abundance.

**Figure S1.** Comparison of divergence estimates using the allele-specific and separate-species analysis pipelines for ribosome protected fragments (RPF), RNA-seq (mRNA) and translational efficiency measurements. The X-axis shows estimates of interspecies differences made by aligning reads to each species genome individually. Estimates of interspecies differences made by allele-specific alignment (Y-axis) are compared for each set.

**Figure S2**. Reproducibility in biological replicates. Scatterplots show comparisons of numbers of normalized reads for each gene from mRNA (top) and RPF (libraries). The value of Spearman's rho is given in the upper left corner of each plot. In all cases, the replicates were highly correlated.

**Figure S3.** Examples of interspecies differences in uORF usage. IGV browser tracks showing normalized ribosome occupancy (RPF) and mRNA abundance (mRNA) in *S. cerevisiae* (blue) and *S. paradoxus* (red). Boxes below depict ORFs (thick) and 5' UTR regions (thin), as well as putative uORFs (small thin lines). (A) *ALR1* mRNA abundance is higher in *S. paradoxus*. However, this species has more ribosome occupancy over uORFs in the 5' UTR, and less ribosome occupancy on the main ORF, compared to *S. cerevisiae*. (B) *S. paradoxus* and *S. cerevisiae* express similar mRNA levels of the uncharacterized ORF YOL014W, but a frameshift mutation in *S. cerevisiae* has resulted in nearly complete loss of ribosome occupancy on its main ORF. Green and black arrows depict translation start and stop sites, respectively.

**Figure S4.** Comparisons of buffered and amplified gene expression levels. Box plots show the distribution of expression levels (in RPKM) in each species for buffered, amplified and all genes.

Figure 1

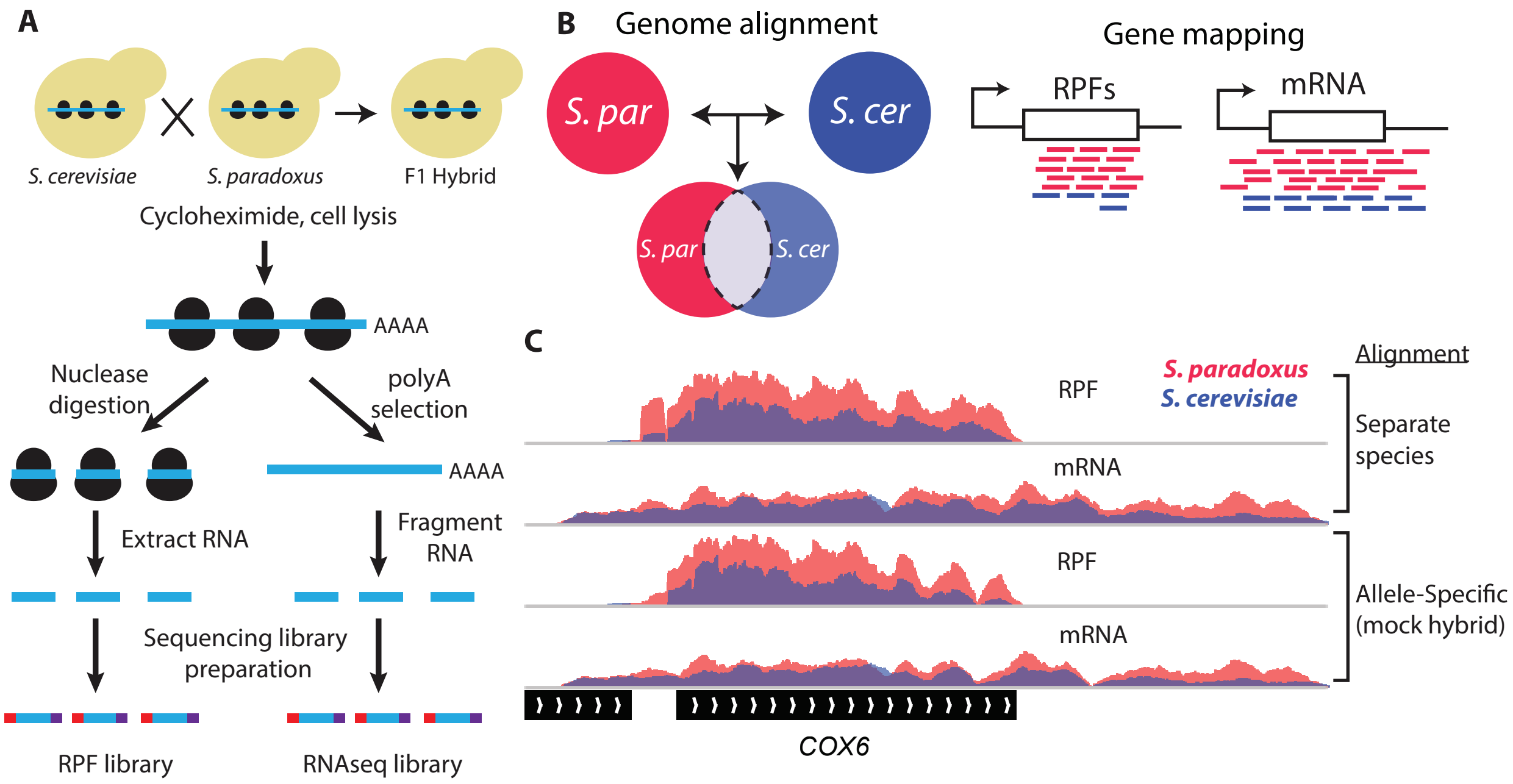

Figure 2

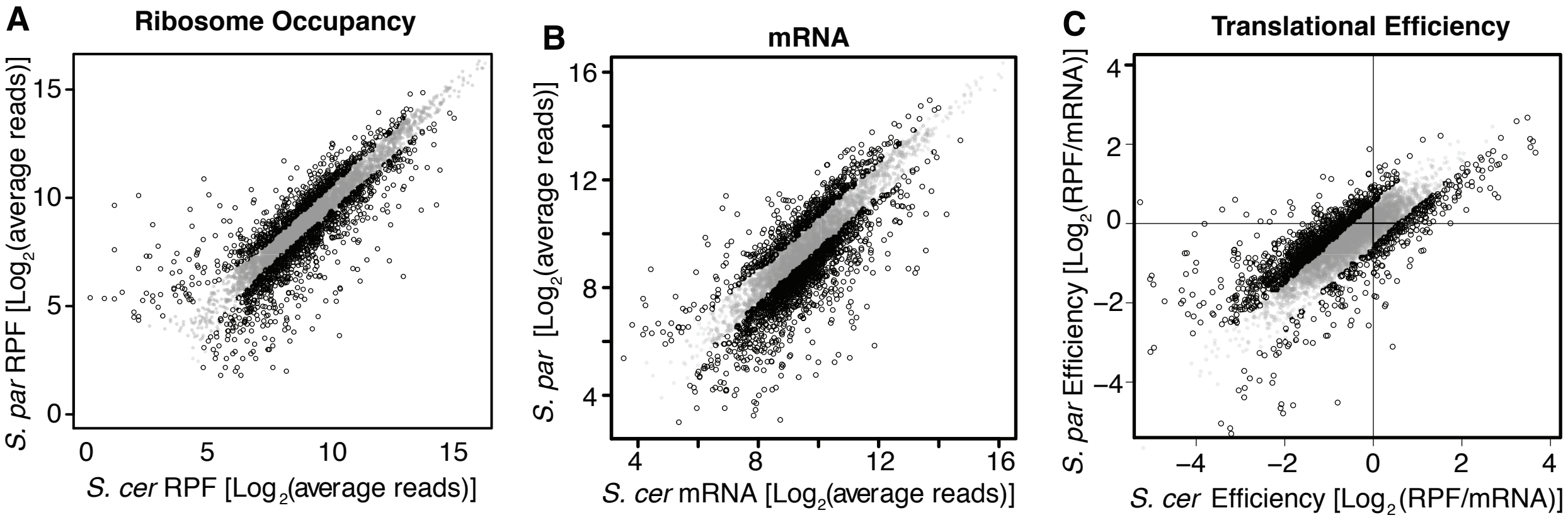
A
Ribosome Occupancy
S. par RPF [Log2(average reads)]
S. cer RPF [Log2(average reads)]
0
5
10
15
B
mRNA
S. par [Log2(average reads)]
S. cer mRNA [Log2(average reads)]
4
8
12
16
C
Translational Efficiency
S. par Efficiency [Log2(RPF/mRNA)]
S. cer Efficiency [Log2(RPF/mRNA)]
−4
−2
0
2
4

Figure 3

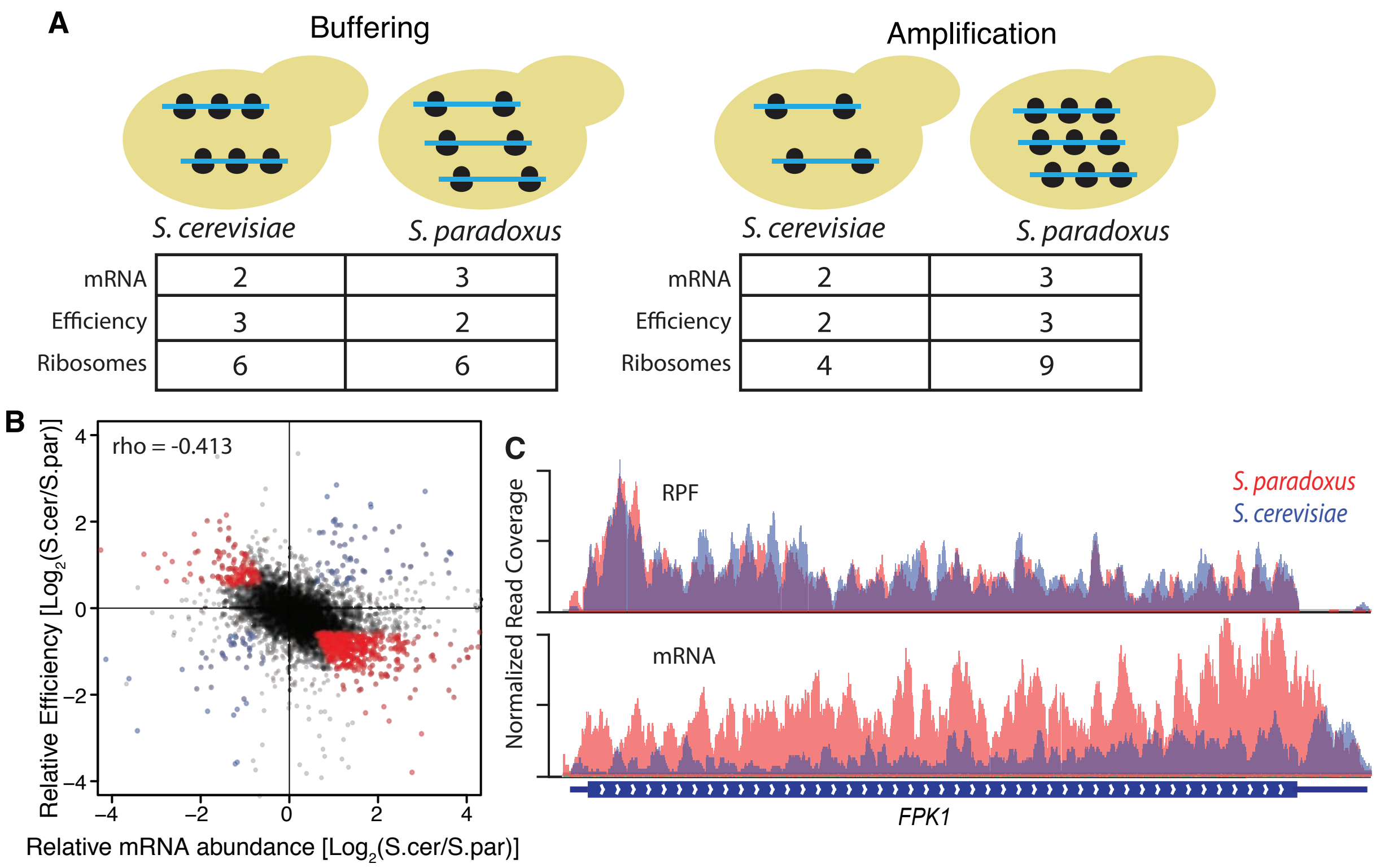

Figure 4

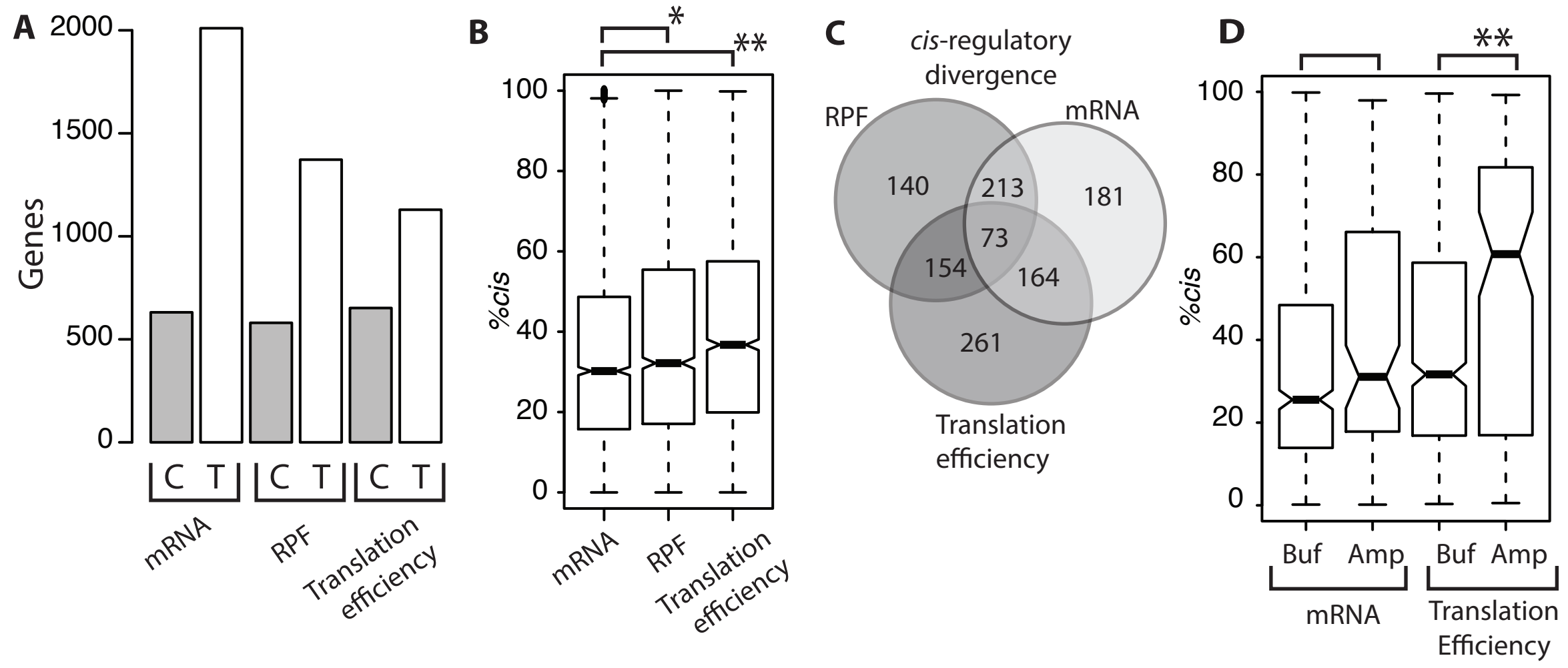

Figure 5

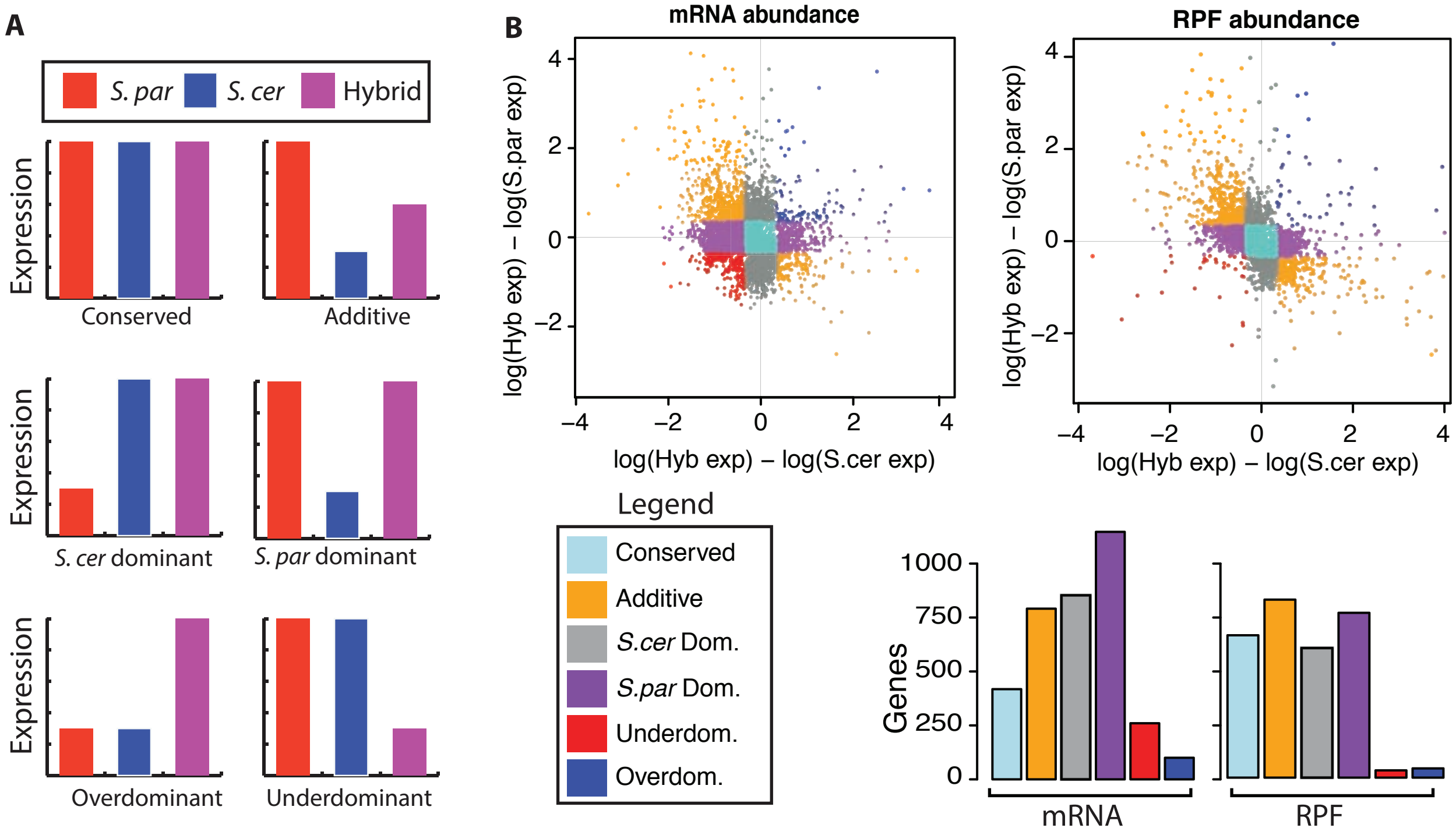

Figure S1

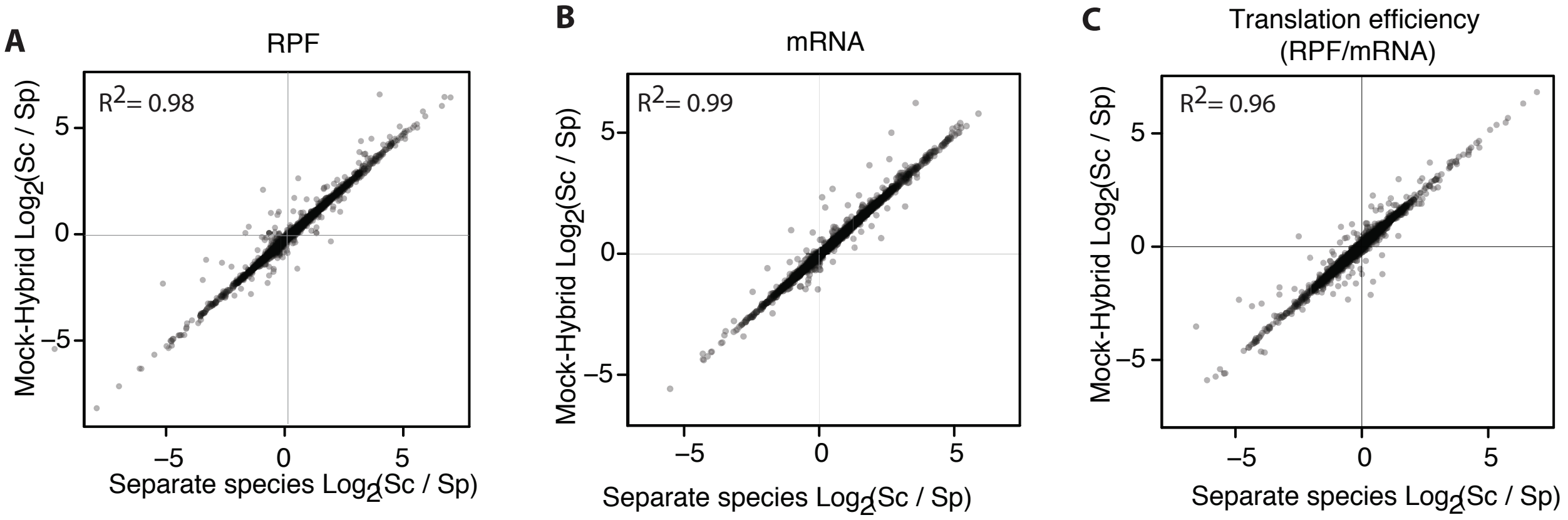
A
RPF
R2= 0.98
Mock-Hybrid Log2(Sc / Sp)
Separate species Log2(Sc / Sp)
B
mRNA
R2= 0.99
Mock-Hybrid Log2(Sc / Sp)
Separate species Log2(Sc / Sp)
C
Translation efficiency
(RPF/mRNA)
R2= 0.96
Mock-Hybrid Log2(Sc / Sp)
Separate species Log2(Sc / Sp)

Figure S2

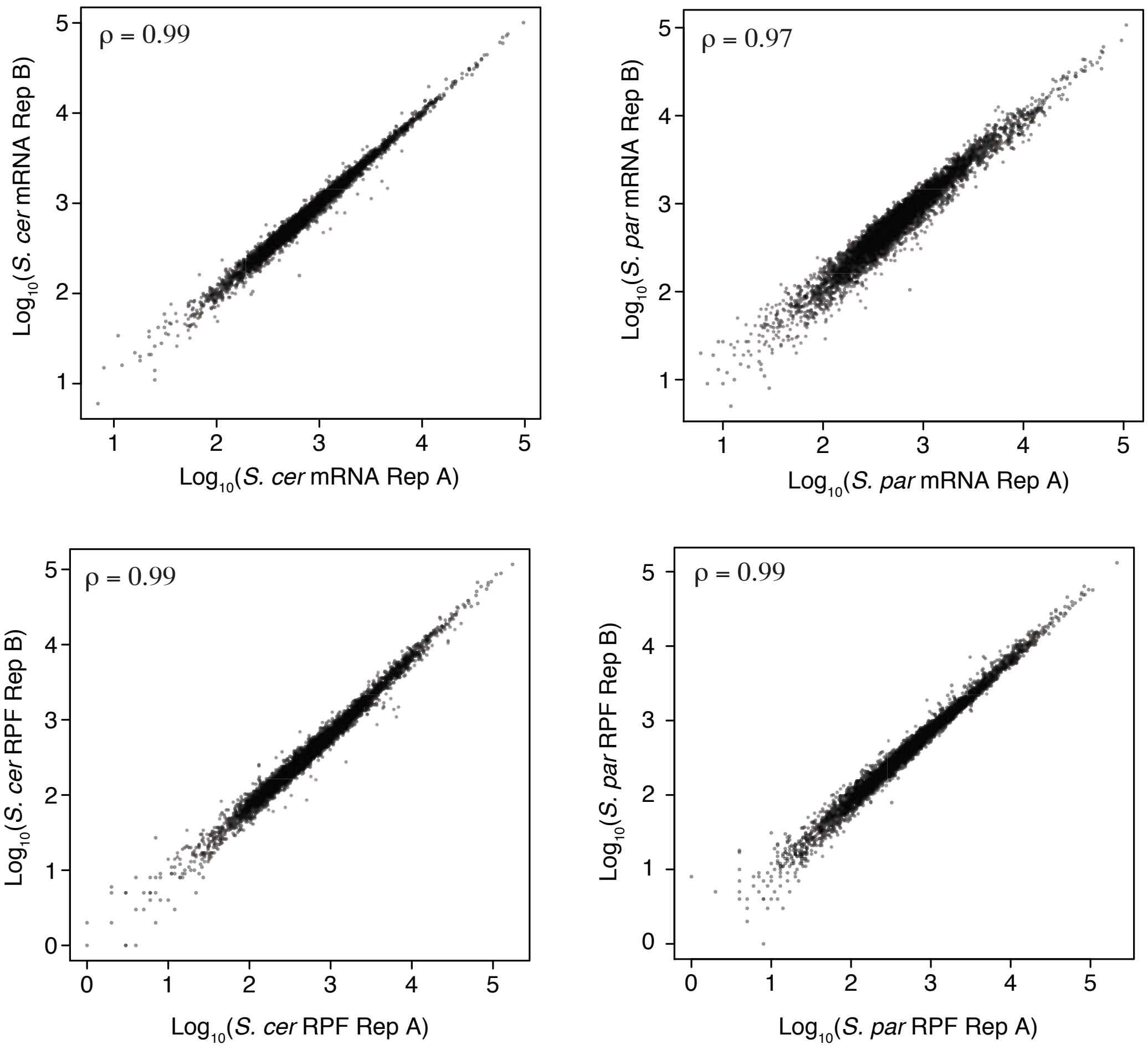

ρ = 0.99
Log10(S. cer mRNA Rep B)
Log10(S. cer mRNA Rep A)
ρ = 0.97
Log10(S. par mRNA Rep B)
Log10(S. par mRNA Rep A)
ρ = 0.99
Log10(S. cer RPF Rep B)
Log10(S. cer RPF Rep A)
ρ = 0.99
Log10(S. par RPF Rep B)
Log10(S. par RPF Rep A)

Figure S3

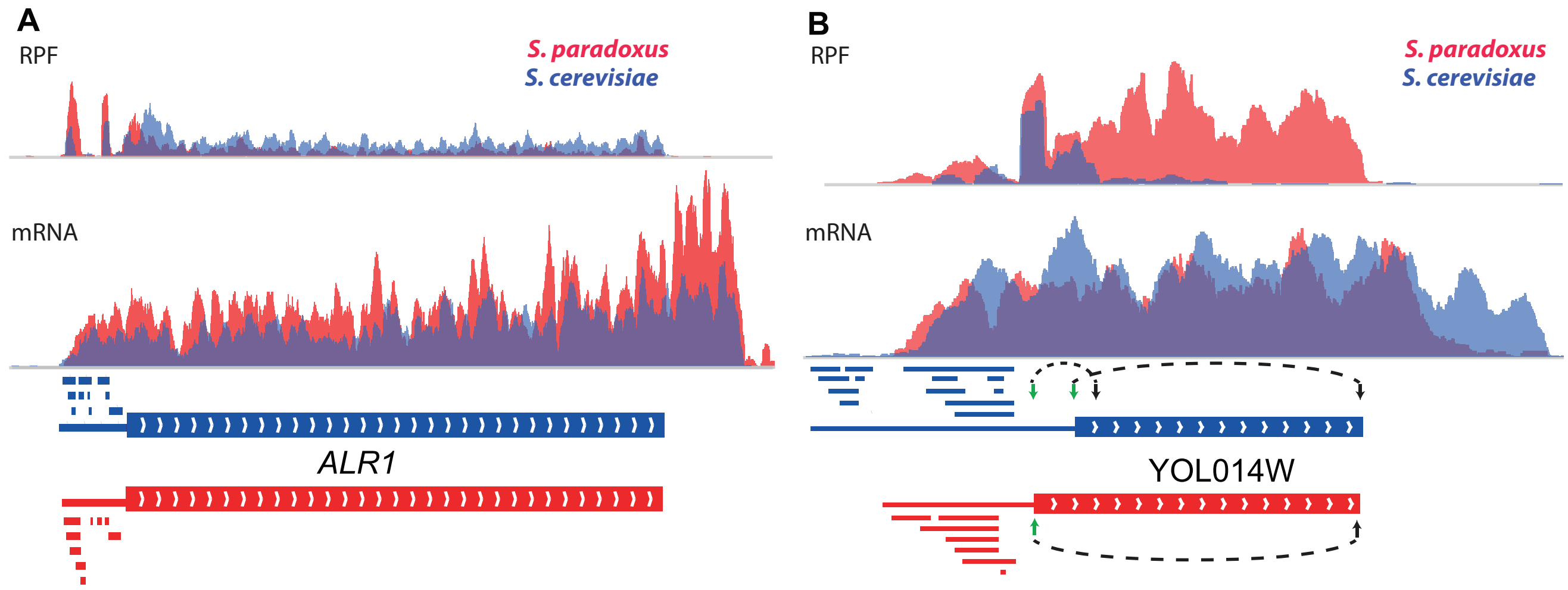

Figure S4

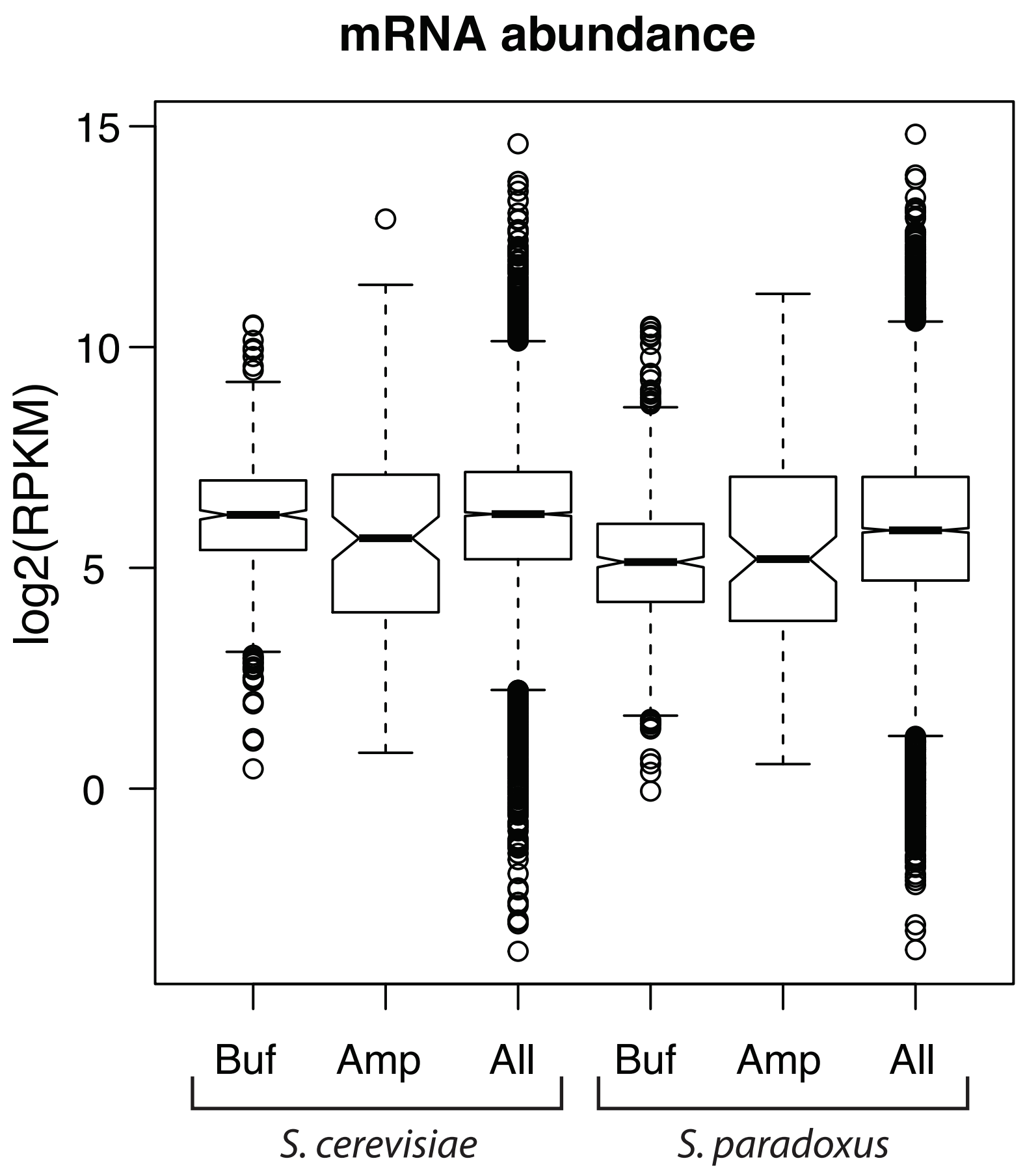

**Table S1.** Numbers of sequence reads at each step of analysis. Raw reads include variable amounts of "adapter only" empty library. Library reads contain cloned RNA inserts. Contaminating rRNA reads made up 30% to 70% of the RPF library reads. As rRNA reads from RPF libraries map to both alleles, the fraction of allele-specific reads is reduced in RPF samples compared to mRNA samples.

| mRNA Sample | Species / Allele | Raw | Library | Genome aligned | Allele-specific | ORF mapped | Normalized | Percent |
|---|---|---|---|---|---|---|---|---|
| Separate Parents Replicate A | *S. paradoxus* | 26,819,693 | 14,160,104 | 13,668,788 | 9,623,380 | 7,195,448 | 7,096,639 | 49.96 |
| | *S. cerevisiae* | 15,288,245 | 13,937,492 | 13,644,258 | 10,085,330 | 7,472,137 | 7,109,313 | 50.04 |
| Separate Parents Replicate B | *S. paradoxus* | 15,222,810 | 13,364,839 | 13,087,419 | 9,434,828 | 7,281,735 | 7,190,496 | 49.96 |
| | *S. cerevisiae* | 15,250,639 | 13,978,236 | 13,670,183 | 10,084,366 | 7,440,143 | 7,201,063 | 50.04 |
| Hybrid Replicate A | *S. paradoxus* | 30,762,220 | 27,877,353 | 26,980,719 | 9,384,780 | 7,011,265 | 6,922,017 | 49.96 |
| | *S. cerevisiae* | | | | 9,657,613 | 7,154,185 | 6,932,791 | 50.04 |
| Hybrid Replicate B | *S. paradoxus* | 27,700,697 | 25,067,868 | 24,382,810 | 8,741,552 | 6,560,924 | 6,454,933 | 49.99 |
| | *S. cerevisiae* | | | | 8,798,257 | 6,599,162 | 6,456,613 | 50.01 |

| RPF Sample | Species / Allele | Raw reads | Library Reads | Aligned reads | Allele-specific | ORF mapped | Normalized | Percent |
|---|---|---|---|---|---|---|---|---|
| Separate Parents Replicate A | *S. paradoxus* | 43,461,865 | 27,640,261 | 26,897,784 | 12,661,113 | 11,291,673 | 8,904,431 | 49.98 |
| | *S. cerevisiae* | 41,214,129 | 22,555,743 | 22,227,255 | 10,526,878 | 9,092,017 | 8,910,405 | 50.02 |
| Separate Parents Replicate B | *S. paradoxus* | 76,161,853 | 36,517,084 | 35,509,272 | 7,247,861 | 6,199,193 | 6,116,133 | 49.95 |
| | *S. cerevisiae* | 38,244,257 | 27,960,534 | 27,568,540 | 10,495,172 | 8,827,413 | 6,127,161 | 50.05 |
| Hybrid Replicate A | *S. paradoxus* | 72,221,925 | 51,421,944 | 50,795,854 | 7,741,869 | 6,879,361 | 6,802,306 | 49.97 |
| | *S. cerevisiae* | | | | 8,374,984 | 7,088,409 | 6,810,286 | 50.03 |
| Hybrid Replicate B | *S. paradoxus* | 97,125,729 | 62,984,843 | 62,255,282 | 8,724,669 | 7,700,918 | 7,615,853 | 49.97 |
| | *S. cerevisiae* | | | | 9,313,643 | 7,894,827 | 7,624,928 | 50.03 |

**Table S2**. Numbers of genes in each category of regulatory divergence using different minimum read thresholds.

| Minimum Read Threshold | Divergent | | | *cis*-acting | | | *trans*-acting | | | Buffered | Amplified |
|---|---|---|---|---|---|---|---|---|---|---|---|
| | RPF | mRNA | Efficiency (RPF/mRNA) | RPF | mRNA | Efficiency (RPF/mRNA) | RPF | mRNA | Efficiency (RPF/mRNA) | | |
| 10 | 1,345 | 1,795 | 1,379 | 601 | 668 | 682 | 1,401 | 2,068 | 1,149 | 574 | 104 |
| 20 | 1,308 | 1,739 | 1,345 | 580 | 631 | 652 | 1,372 | 2,010 | 1,129 | 560 | 101 |
| 50 | 1,237 | 1,649 | 1,279 | 528 | 578 | 591 | 1,309 | 1,909 | 1,099 | 538 | 88 |
| 100 | 1,149 | 1,546 | 1,193 | 457 | 516 | 510 | 1,213 | 1,795 | 1,043 | 500 | 69 |

**Table S3**. Gene ontology slim categories enriched in Buffered and Amplified genes.

| Buffered | | |
|---|---|---|
| term_ID | description | log10 p-value |
| GO:0007154 | cell communication | -2.0023 |
| GO:0009056 | catabolic process | -2.2184 |
| GO:0009987 | cellular process | -2.2339 |
| GO:0044699 | single-organism process | -3.9966 |
| GO:0051179 | localization | -2.6193 |
| GO:0044248 | cellular catabolic process | -2.2489 |
| GO:0044763 | single-organism cellular process | -2.1546 |

| Amplified (higher expression in *S. cerevisiae*) | | |
|---|---|---|
| term_ID | description | log10 p-value |
| GO:1901605 | alpha-amino acid metabolic process | -4.4949 |
| GO:0044710 | single-organism metabolic process | -2.541 |
| GO:0000103 | sulfate assimilation | -2.4692 |
| GO:1901564 | organonitrogen compound metabolic process | -3.3163 |
| GO:0006082 | organic acid metabolic process | -4.5749 |
| GO:0044281 | small molecule metabolic process | -3.1142 |
| GO:1901565 | organonitrogen compound catabolic process | -3.2689 |
| GO:0009069 | serine family amino acid metabolic process | -2.0884 |
| GO:0009066 | aspartate family amino acid metabolic process | -2.7459 |
| GO:0009067 | aspartate family amino acid biosynthetic process | -2.4238 |
| GO:0019752 | carboxylic acid metabolic process | -4.2096 |
| GO:0043436 | oxoacid metabolic process | -4.5935 |
| GO:0006520 | cellular amino acid metabolic process | -4.3143 |

| Amplified (higher expression in *S. paradoxus*) | | |
|---|---|---|
| term_ID | description | log10 p-value |
| GO:0055114 | oxidation-reduction process | -2.4972 |
| GO:1901678 | iron coordination entity transport | -3.3027 |
| GO:0015688 | iron chelate transport | -2.4883 |

**Table S4**. Gene ontology slim categories enriched in genes misexpressed at the mRNA abundance and ribosome occupancy (RPF) levels. RPF underdominant genes were not enriched in any gene functions.

| mRNA Underdominant (underexpressed) | | | |
|---|---|---|---|
| term_ID | description | log10pvalue | representative |
| GO:0000278 | mitotic cell cycle | -9.2487 | mitotic cell cycle |
| GO:0048285 | organelle fission | -4.8245 | mitotic cell cycle |
| GO:0000280 | nuclear division | -5.2541 | mitotic cell cycle |
| GO:0006996 | organelle organization | -4.7547 | mitotic cell cycle |
| GO:0044763 | single-organism cellular process | -6.2454 | mitotic cell cycle |
| GO:0051276 | chromosome organization | -3.8428 | mitotic cell cycle |
| GO:0007049 | cell cycle | -7.4622 | mitotic cell cycle |
| GO:0007059 | chromosome segregation | -3.2387 | mitotic cell cycle |
| GO:0051641 | cellular localization | -2.6559 | mitotic cell cycle |
| GO:0051301 | cell division | -5.1942 | mitotic cell cycle |
| GO:0009057 | macromolecule catabolic process | -3.5642 | macromolecule catabolism |
| GO:0090304 | nucleic acid metabolic process | -7.2549 | macromolecule catabolism |
| GO:0043632 | modification-dependent macromolecule catabolic process | -2.9545 | macromolecule catabolism |
| GO:0006139 | nucleobase-containing compound metabolic process | -5.6396 | macromolecule catabolism |
| GO:0034641 | cellular nitrogen compound metabolic process | -4.2466 | macromolecule catabolism |
| GO:0006366 | transcription from RNA polymerase II promoter | -2.4759 | macromolecule catabolism |
| GO:0016070 | RNA metabolic process | -5.1353 | macromolecule catabolism |
| GO:0006464 | cellular protein modification process | -2.0525 | macromolecule catabolism |
| GO:0044260 | cellular macromolecule metabolic process | -5.4987 | macromolecule catabolism |
| GO:0009987 | cellular process | -5.4116 | cellular process |
| GO:0044699 | single-organism process | -5.2674 | single-organism process |
| GO:0065007 | biological regulation | -6.4058 | biological regulation |
| GO:0071840 | cellular component organization or biogenesis | -6.8034 | cellular component organization or biogenesis |
| GO:0050794 | regulation of cellular process | -6.5348 | regulation of cellular process |
| GO:0043170 | macromolecule metabolic process | -4.7778 | macromolecule metabolism |
| GO:1901360 | organic cyclic compound metabolic process | -3.9786 | macromolecule metabolism |
| GO:0046483 | heterocycle metabolic process | -4.6524 | heterocycle metabolism |
| GO:0006725 | cellular aromatic compound metabolic process | -4.6309 | heterocycle metabolism |
| GO:0006807 | nitrogen compound metabolic process | -2.9407 | nitrogen compound metabolism |

| mRNA Overdominant (overexpressed) | | | |
|---|---|---|---|
| term_ID | description | log10pvalue | representative |
| GO:0006164 | purine nucleotide biosynthetic process | -9.6069 | purine nucleotide biosynthesis |
| GO:1901576 | organic substance biosynthetic process | -2.5486 | purine nucleotide biosynthesis |
| GO:0055086 | nucleobase-containing small molecule metabolic process | -4.9392 | purine nucleotide biosynthesis |
| GO:1901564 | organonitrogen compound metabolic process | -10.6453 | purine nucleotide biosynthesis |
| GO:0046112 | nucleobase biosynthetic process | -2.6326 | purine nucleotide biosynthesis |
| GO:1901566 | organonitrogen compound biosynthetic process | -9.519 | purine nucleotide biosynthesis |
| GO:1901137 | carbohydrate derivative biosynthetic process | -6.1664 | purine nucleotide biosynthesis |
| GO:1901605 | alpha-amino acid metabolic process | -4.5918 | purine nucleotide biosynthesis |
| GO:0009123 | nucleoside monophosphate metabolic process | -7.1679 | purine nucleotide biosynthesis |
| GO:0044281 | small molecule metabolic process | -8.7138 | purine nucleotide biosynthesis |
| GO:0072522 | purine-containing compound biosynthetic process | -8.7296 | purine nucleotide biosynthesis |
| GO:0072521 | purine-containing compound metabolic process | -4.5616 | purine nucleotide biosynthesis |
| GO:0006163 | purine nucleotide metabolic process | -4.4721 | purine nucleotide biosynthesis |
| GO:0090407 | organophosphate biosynthetic process | -4.1909 | purine nucleotide biosynthesis |
| GO:0046394 | carboxylic acid biosynthetic process | -2.5264 | purine nucleotide biosynthesis |
| GO:0019752 | carboxylic acid metabolic process | -3.632 | purine nucleotide biosynthesis |
| GO:0055114 | oxidation-reduction process | -2.8902 | purine nucleotide biosynthesis |
| GO:0006082 | organic acid metabolic process | -3.378 | purine nucleotide biosynthesis |
| GO:0042440 | pigment metabolic process | -3.1396 | purine nucleotide biosynthesis |
| GO:0006811 | ion transport | -2.3434 | ion transport |
| GO:0055085 | transmembrane transport | -2.8476 | ion transport |
| GO:0034220 | ion transmembrane transport | -2.415 | ion transport |
| GO:0008152 | metabolic process | -2.0516 | metabolism |
| GO:0009987 | cellular process | -2.5053 | cellular process |
| GO:0044710 | single-organism metabolic process | -10.6148 | single-organism metabolism |
| GO:1901135 | carbohydrate derivative metabolic process | -4.7286 | carbohydrate derivative metabolism |
| GO:0009058 | biosynthetic process | -2.374 | biosynthesis |
| GO:0006793 | phosphorus metabolic process | -3.1591 | phosphorus metabolism |

| RPF Overdominant (overexpressed) | | | |
|---|---|---|---|
| term_ID | description | log10pvalue | representative |
| GO:0006189 | 'de novo' IMP biosynthetic process | -14.8967 | primede novoprime IMP biosynthesis |
| GO:0055086 | nucleobase-containing small molecule metabolic process | -5.2224 | primede novoprime IMP biosynthesis |
| GO:1901564 | organonitrogen compound metabolic process | -8.0014 | primede novoprime IMP biosynthesis |
| GO:0046112 | nucleobase biosynthetic process | -4.9137 | primede novoprime IMP biosynthesis |
| GO:1901566 | organonitrogen compound biosynthetic process | -8.7202 | primede novoprime IMP biosynthesis |
| GO:1901137 | carbohydrate derivative biosynthetic process | -4.0189 | primede novoprime IMP biosynthesis |
| GO:1901605 | alpha-amino acid metabolic process | -5.7854 | primede novoprime IMP biosynthesis |
| GO:0019438 | aromatic compound biosynthetic process | -2.0551 | primede novoprime IMP biosynthesis |
| GO:0009123 | nucleoside monophosphate metabolic process | -9.6858 | primede novoprime IMP biosynthesis |
| GO:0044281 | small molecule metabolic process | -7.3388 | primede novoprime IMP biosynthesis |
| GO:0072522 | purine-containing compound biosynthetic process | -12.6204 | primede novoprime IMP biosynthesis |
| GO:0072521 | purine-containing compound metabolic process | -7.3924 | primede novoprime IMP biosynthesis |
| GO:0006163 | purine nucleotide metabolic process | -5.6913 | primede novoprime IMP biosynthesis |
| GO:0090407 | organophosphate biosynthetic process | -3.8273 | primede novoprime IMP biosynthesis |
| GO:0019878 | lysine biosynthetic process via aminoadipic acid | -2.3469 | primede novoprime IMP biosynthesis |
| GO:0019752 | carboxylic acid metabolic process | -5.2161 | primede novoprime IMP biosynthesis |
| GO:0006082 | organic acid metabolic process | -5.0142 | primede novoprime IMP biosynthesis |
| GO:0009066 | aspartate family amino acid metabolic process | -3.3552 | primede novoprime IMP biosynthesis |
| GO:0042440 | pigment metabolic process | -5.8078 | primede novoprime IMP biosynthesis |
| GO:0044710 | single-organism metabolic process | -7.1985 | single-organism metabolism |
| GO:1901135 | carbohydrate derivative metabolic process | -2.4313 | carbohydrate derivative metabolism |

**Table S5.** Overview of normalized readcounts and results of statistical tests for divergent expression, and *cis*- and *trans*-regulatory divergence. "Eff" denotes translation efficiency (RPF / mRNA). Percent cis denotes the percentage of regulatory divergence attributable to *cis*-regulatory differences. "Padj" denotes p-values adjusted for FDR correction using the Benjamini-Hothberg method.

| | Mixed Parental RPF (ribosome occupancy) read counts | | | | Mixed Parental mRNA read counts | | | | F1 hybrid RPF (ribosome occupancy) read counts | | | | F1 hybrid mRNA read counts | | | | Ratios between Species | | | Ratios between Alleles in F1 hybrid | | | Percent *cis* | | |
|---|---|---|---|---|---|---|---|---|---|---|---|---|---|---|---|---|---|---|---|---|---|---|---|---|---|
| | *S. cerevisiae* | | *S. paradoxus* | | *S. cerevisiae* | | *S. paradoxus* | | *S. cerevisiae* | | *S. paradoxus* | | *S. cerevisiae* | | *S. paradoxus* | | | | | | | | | | |
| Gene | Rep A | Rep B | Rep A | Rep B | Rep A | Rep B | Rep A | Rep B | Rep A | Rep B | Rep A | Rep B | Rep A | Rep B | Rep A | Rep B | RPF | mRNA | Eff | RPF | mRNA | Eff | RPF | mRNA | Eff |
| YKR046C | 428 | 317 | 2948 | 2421 | 356 | 338 | 2155 | 2210 | 702 | 875 | 1684 | 2324 | 750 | 1203 | 1949 | 2673 | 0.14 | 0.16 | 0.87 | 0.39 | 0.42 | 0.93 | 47.23 | 46.85 | 50.02 |
| YHR071W | 99 | 49 | 339 | 256 | 367 | 274 | 2545 | 2452 | 37 | 48 | 103 | 83 | 411 | 243 | 1443 | 985 | 0.25 | 0.13 | 1.91 | 0.46 | 0.27 | 1.71 | 56.28 | 63.88 | 82.93 |
| YOR317W | 5069 | 3660 | 833 | 571 | 4706 | 5405 | 455 | 769 | 1445 | 1755 | 811 | 1203 | 1752 | 3053 | 764 | 1727 | 6.22 | 8.26 | 0.74 | 1.59 | 1.93 | 0.81 | 25.34 | 31.11 | 70.05 |
| YIL051C | 8014 | 5614 | 3103 | 2479 | 6654 | 7255 | 2192 | 1983 | 4204 | 4684 | 5154 | 6220 | 5318 | 3893 | 4687 | 4160 | 2.44 | 3.33 | 0.73 | 0.78 | 1.04 | 0.76 | 17.80 | 3.35 | 89.19 |
| YDR481C | 4235 | 2769 | 660 | 409 | 4617 | 4653 | 925 | 816 | 959 | 1028 | 982 | 887 | 1681 | 2939 | 2010 | 3243 | 6.55 | 5.32 | 1.24 | 1.06 | 0.88 | 1.23 | 3.26 | 6.66 | 93.51 |
| YOR273C | 946 | 962 | 711 | 519 | 3151 | 3811 | 495 | 560 | 769 | 1866 | 292 | 479 | 1975 | 4700 | 349 | 669 | 1.55 | 6.60 | 0.24 | 3.42 | 6.56 | 0.52 | 60.87 | 99.66 | 45.22 |
| YDR533C | 791 | 917 | 43 | 33 | 1255 | 1990 | 35 | 56 | 181 | 263 | 35 | 37 | 338 | 453 | 68 | 71 | 22.47 | 35.66 | 0.64 | 6.17 | 5.69 | 1.08 | 58.45 | 48.65 | 12.57 |
| YPL126W | 2693 | 1333 | 3276 | 2154 | 1760 | 1545 | 2951 | 4695 | 2198 | 2002 | 1924 | 1777 | 2174 | 1388 | 2069 | 1300 | 0.74 | 0.43 | 1.57 | 1.13 | 1.06 | 1.07 | 22.91 | 5.86 | 15.63 |
| YMR058W | 5996 | 1369 | 8108 | 5506 | 1768 | 780 | 3063 | 3393 | 3917 | 3409 | 7483 | 7404 | 1933 | 2780 | 3302 | 5426 | 0.54 | 0.39 | 1.21 | 0.49 | 0.54 | 0.90 | 88.22 | 66.28 | 26.50 |
| YLR214W | 6827 | 1471 | 4614 | 4310 | 4601 | 1461 | 2827 | 5128 | 2491 | 3026 | 3548 | 4090 | 2368 | 2019 | 2930 | 3601 | 0.93 | 0.76 | 1.00 | 0.72 | 0.67 | 1.08 | 56.29 | 75.93 | 49.70 |
| YFR055W | 1636 | 708 | 4653 | 3635 | 1480 | 786 | 3115 | 4183 | 1300 | 1816 | 2018 | 2629 | 1451 | 1811 | 2230 | 3367 | 0.28 | 0.31 | 0.84 | 0.67 | 0.58 | 1.15 | 31.65 | 46.16 | 30.50 |
| YMR290C | 2317 | 1229 | 4355 | 2770 | 1512 | 1308 | 3215 | 4667 | 2258 | 2207 | 2645 | 2642 | 1744 | 1104 | 2869 | 1351 | 0.50 | 0.36 | 1.31 | 0.84 | 0.67 | 1.23 | 24.22 | 38.26 | 78.15 |
| YMR305C | 3167 | 2016 | 5480 | 3525 | 2230 | 2288 | 4451 | 5732 | 2090 | 2396 | 2576 | 2865 | 2110 | 3374 | 2638 | 4331 | 0.58 | 0.44 | 1.27 | 0.82 | 0.79 | 1.05 | 34.94 | 29.49 | 19.31 |
| YJL198W | 1219 | 766 | 1962 | 1287 | 1854 | 1815 | 3988 | 5024 | 1140 | 1014 | 1251 | 1196 | 2221 | 3301 | 3238 | 4134 | 0.61 | 0.41 | 1.46 | 0.88 | 0.75 | 1.19 | 25.88 | 32.15 | 45.03 |
| YBR084W | 4219 | 2466 | 5803 | 3882 | 1955 | 2109 | 4257 | 5696 | 3885 | 3624 | 3940 | 3805 | 3329 | 3208 | 4140 | 6308 | 0.69 | 0.41 | 1.64 | 0.97 | 0.63 | 1.52 | 8.35 | 52.35 | 84.50 |
| YJL196C | 1848 | 1691 | 1415 | 1048 | 2361 | 2676 | 741 | 463 | 862 | 967 | 764 | 834 | 1337 | 1623 | 505 | 511 | 1.44 | 4.18 | 0.35 | 1.14 | 2.91 | 0.39 | 37.25 | 74.72 | 88.34 |
| YGR175C | 4733 | 3325 | 8951 | 5565 | 2135 | 2076 | 4609 | 6028 | 5962 | 6705 | 5409 | 6269 | 3864 | 5374 | 3875 | 5381 | 0.56 | 0.40 | 1.39 | 1.08 | 1.00 | 1.09 | 10.82 | 0.21 | 24.97 |
| YOR374W | 3551 | 2119 | 651 | 428 | 1734 | 2268 | 382 | 352 | 1772 | 1714 | 1019 | 1043 | 923 | 2026 | 669 | 1602 | 5.25 | 5.45 | 0.98 | 1.69 | 1.30 | 1.28 | 31.65 | 15.40 | 48.02 |
| YBR067C | 2949 | 3235 | 752 | 758 | 2233 | 2890 | 593 | 554 | 707 | 693 | 725 | 671 | 700 | 557 | 777 | 486 | 4.10 | 4.47 | 0.92 | 1.00 | 1.00 | 1.00 | 0.20 | 0.32 | 3.63 |
| YGL037C | 3965 | 2808 | 1342 | 1191 | 3007 | 3622 | 1013 | 1014 | 1286 | 2270 | 1272 | 2066 | 1342 | 1758 | 1277 | 1349 | 2.67 | 3.27 | 0.82 | 1.07 | 1.18 | 0.89 | 6.43 | 14.00 | 56.60 |
| YGR088W | 142 | 136 | 98 | 47 | 1496 | 1950 | 208 | 274 | 60 | 156 | 109 | 192 | 555 | 1589 | 445 | 1385 | 1.92 | 7.15 | 0.27 | 0.72 | 1.17 | 0.61 | 25.24 | 8.05 | 37.53 |
| YMR006C | 1968 | 1324 | 5747 | 4115 | 1094 | 1237 | 3298 | 4292 | 2686 | 2284 | 3512 | 4143 | 1856 | 2747 | 3306 | 6499 | 0.33 | 0.31 | 1.07 | 0.65 | 0.47 | 1.33 | 39.37 | 64.05 | 56.94 |
| YDL039C | 8 | 4 | 508 | 362 | 217 | 98 | 1670 | 1719 | 31 | 39 | 191 | 62 | 408 | 913 | 249 | 414 | 0.01 | 0.09 | 0.14 | 0.28 | 1.99 | 0.16 | 30.00 | 18.36 | 94.37 |
| YNL111C | 976 | 610 | 3956 | 3428 | 478 | 434 | 2334 | 2560 | 2479 | 2363 | 2501 | 2388 | 1627 | 1283 | 1594 | 1282 | 0.21 | 0.19 | 1.13 | 0.99 | 1.01 | 0.98 | 0.63 | 0.69 | 12.62 |
| YMR145C | 3058 | 1981 | 7565 | 4432 | 1987 | 1764 | 5036 | 4889 | 4061 | 4739 | 4093 | 4829 | 4332 | 4051 | 4210 | 4171 | 0.42 | 0.38 | 1.11 | 0.99 | 1.00 | 0.99 | 1.59 | 0.02 | 9.37 |
| YHR143W | 2848 | 2261 | 631 | 606 | 2516 | 2462 | 541 | 414 | 1217 | 1248 | 977 | 1152 | 1343 | 1976 | 669 | 879 | 4.13 | 5.21 | 0.79 | 1.16 | 2.14 | 0.54 | 10.33 | 46.19 | 61.57 |
| YML075C | 1294 | 802 | 6652 | 4647 | 1233 | 1250 | 5121 | 6875 | 2634 | 2561 | 4477 | 4223 | 2831 | 4121 | 4198 | 6421 | 0.19 | 0.21 | 0.87 | 0.60 | 0.65 | 0.91 | 30.61 | 26.89 | 66.22 |
| YBR092C | 2438 | 1799 | 110 | 98 | 2107 | 2098 | 229 | 267 | 1631 | 1914 | 497 | 434 | 1513 | 2268 | 1534 | 3892 | 20.37 | 8.48 | 2.37 | 3.81 | 0.70 | 5.21 | 44.36 | 12.63 | 67.73 |
| YHR136C | 1435 | 1090 | 14 | 11 | 2090 | 2125 | 54 | 30 | 655 | 872 | 645 | 968 | 1161 | 1448 | 1104 | 1488 | 101.00 | 50.18 | 2.06 | 0.95 | 1.01 | 0.94 | 1.16 | 0.17 | 7.04 |
| YML100W | 2141 | 1565 | 142 | 72 | 2419 | 2497 | 101 | 136 | 440 | 843 | 344 | 726 | 598 | 2243 | 475 | 2709 | 17.32 | 20.74 | 0.83 | 1.20 | 0.89 | 1.28 | 6.37 | 3.50 | 36.73 |
| YMR121C | 1205 | 847 | 7027 | 4747 | 1272 | 1246 | 7006 | 5098 | 1103 | 1266 | 4929 | 5836 | 1508 | 1073 | 5707 | 3730 | 0.17 | 0.21 | 0.84 | 0.22 | 0.27 | 0.80 | 86.65 | 82.57 | 81.92 |
| YNL036W | 6431 | 1088 | 323 | 79 | 6347 | 3798 | 735 | 105 | 1044 | 5279 | 1111 | 4741 | 1685 | 10071 | 1206 | 8479 | 18.70 | 12.08 | 1.76 | 1.08 | 1.21 | 0.89 | 2.64 | 7.78 | 14.70 |
| YIL094C | 2593 | 1840 | 307 | 220 | 2536 | 2418 | 432 | 348 | 3294 | 3170 | 915 | 826 | 3517 | 2174 | 1110 | 733 | 8.41 | 6.35 | 1.33 | 3.71 | 3.09 | 1.20 | 61.60 | 60.99 | 64.14 |
| YDR001C | 1558 | 1035 | 71 | 40 | 2290 | 2315 | 78 | 104 | 482 | 741 | 139 | 152 | 1038 | 1184 | 148 | 285 | 23.36 | 25.30 | 0.91 | 4.20 | 5.13 | 0.83 | 45.56 | 50.62 | 66.08 |
| YDL192W | 5623 | 4367 | 1455 | 1309 | 4265 | 3992 | 1414 | 1229 | 4295 | 4826 | 1234 | 1497 | 3951 | 1851 | 1329 | 1012 | 3.61 | 3.12 | 1.16 | 3.34 | 2.48 | 1.41 | 93.85 | 79.68 | 63.35 |
| YDL124W | 2862 | 2027 | 615 | 390 | 3296 | 3579 | 686 | 657 | 581 | 776 | 390 | 476 | 888 | 1415 | 786 | 1187 | 4.86 | 5.12 | 0.96 | 1.57 | 1.17 | 1.35 | 28.39 | 9.47 | 46.96 |
| YER072W | 8549 | 7010 | 1596 | 1099 | 6839 | 6515 | 1566 | 1171 | 3714 | 3707 | 3279 | 3318 | 3828 | 3482 | 3436 | 2180 | 5.77 | 4.88 | 1.19 | 1.12 | 1.30 | 0.86 | 6.71 | 16.63 | 32.02 |
| YNR060W | 35 | 17 | 2319 | 1451 | 127 | 116 | 2392 | 3029 | 379 | 208 | 509 | 217 | 546 | 494 | 730 | 529 | 0.01 | 0.04 | 0.29 | 0.81 | 0.83 | 1.01 | 4.96 | 6.15 | 0.55 |
| YDR074W | 2721 | 2015 | 1447 | 901 | 5167 | 5512 | 805 | 1210 | 1140 | 2113 | 1409 | 2468 | 1624 | 5042 | 2263 | 5635 | 2.02 | 5.30 | 0.37 | 0.84 | 0.84 | 1.01 | 16.67 | 8.45 | 0.86 |
| YKL035W | 5468 | 4077 | 3072 | 2033 | 5621 | 7072 | 1910 | 2274 | 2616 | 3621 | 2570 | 3418 | 3113 | 5554 | 2705 | 4372 | 1.87 | 3.03 | 0.62 | 1.04 | 1.22 | 0.85 | 6.51 | 18.26 | 33.43 |
| YMR195W | 2252 | 1788 | 331 | 393 | 2717 | 3391 | 381 | 462 | 843 | 1405 | 863 | 1530 | 1250 | 1364 | 1351 | 1327 | 5.58 | 7.25 | 0.76 | 0.94 | 0.98 | 0.96 | 3.39 | 1.19 | 15.27 |
| YMR250W | 714 | 550 | 84 | 84 | 2002 | 2341 | 107 | 162 | 104 | 233 | 100 | 277 | 505 | 1697 | 433 | 1328 | 7.52 | 16.14 | 0.45 | 0.89 | 1.25 | 0.72 | 5.00 | 8.03 | 42.18 |
| YMR303C | 8327 | 7433 | 73 | 93 | 3590 | 3858 | 48 | 67 | 5990 | 6832 | 64 | 58 | 3443 | 3271 | 74 | 109 | 94.94 | 64.77 | 1.44 | 105.10 | 36.69 | 2.86 | 97.86 | 86.37 | 60.58 |
| YGR213C | 27 | 8 | 730 | 691 | 134 | 90 | 2204 | 1741 | 20 | 8 | 414 | 431 | 110 | 148 | 1689 | 2436 | 0.02 | 0.06 | 0.44 | 0.03 | 0.06 | 0.52 | 91.99 | 96.63 | 79.71 |
| YOR161C | 754 | 689 | 99 | 72 | 2041 | 2367 | 131 | 82 | 263 | 357 | 115 | 258 | 638 | 1705 | 218 | 748 | 8.44 | 20.69 | 0.42 | 1.66 | 2.43 | 0.66 | 23.82 | 29.24 | 48.11 |
| YAR015W | 871 | 586 | 1824 | 2970 | 1052 | 954 | 4155 | 5744 | 1590 | 4975 | 2425 | 6422 | 3203 | 4224 | 5137 | 5638 | 0.30 | 0.20 | 1.51 | 0.74 | 0.69 | 1.04 | 25.05 | 23.31 | 9.40 |
| YDL022W | 3977 | 3547 | 1978 | 1378 | 3164 | 3669 | 570 | 686 | 2284 | 3345 | 1199 | 1935 | 1444 | 2546 | 864 | 1509 | 2.24 | 5.44 | 0.42 | 1.80 | 1.68 | 1.07 | 72.54 | 30.68 | 6.39 |
| YLR258W | 1316 | 937 | 49 | 49 | 2870 | 3387 | 152 | 162 | 119 | 232 | 140 | 298 | 320 | 1136 | 435 | 1306 | 22.99 | 19.93 | 1.15 | 0.80 | 0.84 | 0.97 | 6.19 | 5.34 | 15.83 |
| YHL021C | 1796 | 1513 | 435 | 353 | 3306 | 3877 | 318 | 432 | 639 | 1610 | 537 | 1295 | 1168 | 3712 | 1016 | 3174 | 4.20 | 9.58 | 0.44 | 1.23 | 1.16 | 1.06 | 14.29 | 6.75 | 5.73 |
| YJL052W | 7572 | 8159 | 19284 | 12603 | 4884 | 6012 | 13539 | 10862 | 2848 | 3378 | 12388 | 14088 | 2432 | 2364 | 12023 | 10731 | 0.49 | 0.45 | 1.13 | 0.24 | 0.21 | 1.11 | 66.14 | 67.47 | 87.61 |
| YJR148W | 6264 | 4477 | 3677 | 2700 | 5393 | 5999 | 2035 | 2569 | 4568 | 5611 | 1588 | 2123 | 4560 | 6290 | 1780 | 2407 | 1.68 | 2.47 | 0.67 | 2.74 | 2.59 | 1.06 | 67.42 | 95.37 | 11.11 |
| YNL160W | 1297 | 1267 | 60 | 55 | 1627 | 1807 | 200 | 224 | 286 | 281 | 251 | 334 | 229 | 376 | 458 | 906 | 22.30 | 8.10 | 2.75 | 0.97 | 0.44 | 2.13 | 0.99 | 21.87 | 74.69 |
| YLR355C | 9652 | 7534 | 22048 | 13235 | 3510 | 3368 | 6152 | 7157 | 13752 | 15666 | 11491 | 13282 | 4552 | 5915 | 4195 | 5754 | 0.49 | 0.52 | 0.96 | 1.19 | 1.05 | 1.13 | 16.17 | 6.66 | 42.50 |
| YBR052C | 1285 | 1112 | 149 | 114 | 1288 | 1433 | 108 | 98 | 625 | 813 | 232 | 275 | 796 | 823 | 195 | 208 | 9.11 | 13.21 | 0.70 | 2.84 | 4.02 | 0.71 | 47.18 | 53.88 | 96.83 |
| YGR155W | 15130 | 10081 | 15514 | 9421 | 6586 | 6624 | 9113 | 11602 | 12776 | 12805 | 12496 | 12069 | 8676 | 7397 | 10601 | 8494 | 1.01 | 0.64 | 1.57 | 1.04 | 0.84 | 1.23 | 57.86 | 38.30 | 46.44 |
| YMR307W | 26578 | 18614 | 25290 | 17726 | 8801 | 9396 | 14397 | 15175 | 18238 | 20142 | 18755 | 20985 | 10730 | 12977 | 11349 | 14192 | 1.05 | 0.62 | 1.71 | 0.97 | 0.93 | 1.04 | 29.26 | 15.35 | 7.32 |
| YLR409C | 1253 | 825 | 2392 | 1524 | 1500 | 1408 | 2936 | 4083 | 1422 | 1266 | 1501 | 1395 | 1923 | 1678 | 2482 | 3326 | 0.53 | 0.41 | 1.23 | 0.93 | 0.62 | 1.47 | 11.76 | 54.25 | 68.47 |
| YDR232W | 2152 | 1569 | 3957 | 2349 | 550 | 591 | 2167 | 2579 | 2505 | 2577 | 2903 | 2676 | 1086 | 1764 | 1885 | 2448 | 0.59 | 0.24 | 2.48 | 0.91 | 0.66 | 1.41 | 17.69 | 29.39 | 37.43 |
| YMR011W | 4508 | 1676 | 9802 | 6386 | 595 | 426 | 2576 | 2224 | 13623 | 12813 | 7556 | 6229 | 3383 | 4474 | 1453 | 1887 | 0.38 | 0.21 | 1.72 | 1.92 | 2.35 | 0.82 | 28.75 | 26.25 | 20.83 |
| YDR055W | 2819 | 2065 | 1837 | 1706 | 6679 | 6945 | 2026 | 3310 | 1401 | 1785 | 1090 | 1551 | 3695 | 5545 | 2155 | 3820 | 1.38 | 2.55 | 0.52 | 1.21 | 1.55 | 0.77 | 58.45 | 46.51 | 38.88 |
| YBR177C | 1847 | 1340 | 528 | 408 | 2264 | 2554 | 454 | 593 | 837 | 1210 | 1074 | 1286 | 1189 | 2070 | 1380 | 2253 | 3.40 | 4.60 | 0.73 | 0.87 | 0.90 | 0.97 | 9.42 | 6.23 | 8.51 |
| YJL079C | 1047 | 1261 | 41 | 34 | 848 | 1195 | 27 | 48 | 164 | 304 | 103 | 162 | 177 | 487 | 120 | 216 | 30.77 | 27.24 | 1.13 | 1.77 | 1.98 | 0.91 | 16.60 | 20.61 | 29.64 |
| YNL190W | 3594 | 2516 | 4380 | 3311 | 958 | 1020 | 2512 | 3328 | 3024 | 3422 | 3327 | 3561 | 1091 | 1704 | 3099 | 3669 | 0.79 | 0.34 | 2.31 | 0.94 | 0.41 | 2.28 | 28.82 | 81.69 | 98.60 |
| YBL003C | 5151 | 4002 | 2283 | 1576 | 6627 | 5673 | 4155 | 2476 | 3629 | 3825 | 2861 | 3092 | 4715 | 3002 | 4031 | 2323 | 2.37 | 1.85 | 1.28 | 1.25 | 1.21 | 1.03 | 26.03 | 31.45 | 10.30 |
| YDR225W | 6874 | 5827 | 4530 | 3148 | 10306 | 9423 | 5734 | 5126 | 5493 | 6025 | 5583 | 6033 | 6702 | 7184 | 6980 | 7960 | 1.65 | 1.82 | 0.91 | 0.99 | 0.93 | 1.07 | 1.63 | 9.84 | 29.40 |
| YBR026C | 1291 | 918 | 207 | 158 | 1479 | 1649 | 177 | 204 | 606 | 687 | 236 | 340 | 720 | 627 | 313 | 265 | 6.05 | 8.21 | 0.73 | 2.24 | 2.33 | 0.97 | 44.91 | 40.19 | 9.21 |
| YOR264W | 882 | 603 | 301 | 257 | 2752 | 2765 | 638 | 720 | 449 | 536 | 296 | 223 | 1206 | 773 | 701 | 514 | 2.66 | 4.06 | 0.65 | 1.90 | 1.63 | 1.16 | 65.46 | 34.80 | 20.09 |
| YOR128C | 1523 | 1140 | 1844 | 2442 | 1695 | 1804 | 2563 | 4619 | 2008 | 5863 | 2451 | 6671 | 5565 | 6287 | 5002 | 4976 | 0.62 | 0.49 | 1.22 | 0.86 | 1.19 | 0.71 | 30.99 | 16.19 | 38.73 |
| YMR205C | 20586 | 12955 | 18510 | 10863 | 2674 | 2661 | 5290 | 5920 | 13563 | 15674 | 13716 | 16525 | 4114 | 7410 | 4648 | 8276 | 1.14 | 0.48 | 2.41 | 0.97 | 0.89 | 1.08 | 16.86 | 15.44 | 8.90 |
| YLR134W | 4415 | 2788 | 6418 | 5162 | 1593 | 1507 | 3699 | 4129 | 3516 | 4690 | 5308 | 6213 | 2203 | 1725 | 4598 | 4618 | 0.62 | 0.40 | 1.54 | 0.71 | 0.43 | 1.67 | 71.47 | 92.07 | 86.35 |
| YGL009C | 3232 | 1796 | 4461 | 3512 | 915 | 975 | 1715 | 3218 | 2387 | 2844 | 2680 | 3824 | 1097 | 1420 | 1974 | 2516 | 0.63 | 0.38 | 1.52 | 0.80 | 0.56 | 1.44 | 47.24 | 60.33 | 87.01 |
| YHR006W | 546 | 378 | 331 | 270 | 2083 | 2231 | 478 | 464 | 320 | 328 | 312 | 436 | 861 | 932 | 498 | 485 | 1.54 | 4.58 | 0.34 | 0.87 | 1.82 | 0.48 | 20.01 | 39.50 | 68.24 |
| YGR060W | 6339 | 4451 | 10121 | 7538 | 3416 | 3158 | 6469 | 6618 | 6651 | 7261 | 6824 | 7112 | 5588 | 3832 | 5419 | 4514 | 0.61 | 0.50 | 1.21 | 1.00 | 0.95 | 1.06 | 0.35 | 7.70 | 29.25 |
| YDL131W | 2442 | 1503 | 543 | 380 | 2697 | 2795 | 662 | 754 | 2604 | 2784 | 2048 | 2067 | 3204 | 3646 | 2334 | 2899 | 4.27 | 3.88 | 1.09 | 1.31 | 1.31 | 1.00 | 18.56 | 19.87 | 0.54 |
| YDL063C | 922 | 441 | 2309 | 1578 | 1545 | 1304 | 2836 | 3766 | 1087 | 774 | 1358 | 1055 | 1825 | 793 | 2351 | 1258 | 0.35 | 0.43 | 0.76 | 0.77 | 0.73 | 1.08 | 24.79 | 38.20 | 17.99 |
| YMR229C | 8317 | 3911 | 10825 | 5735 | 1620 | 1182 | 4876 | 3552 | 6661 | 7086 | 6863 | 7229 | 2470 | 1179 | 3082 | 1311 | 0.74 | 0.33 | 2.21 | 0.98 | 0.83 | 1.17 | 8.17 | 16.85 | 19.53 |
| YJL109C | 3984 | 1818 | 5965 | 3381 | 1737 | 1495 | 4127 | 4074 | 2943 | 3030 | 3284 | 3506 | 2720 | 2054 | 2805 | 2696 | 0.62 | 0.39 | 1.54 | 0.88 | 0.87 | 1.02 | 26.89 | 15.22 | 4.63 |
| YDR502C | 4075 | 2669 | 6409 | 4481 | 1432 | 1557 | 3479 | 4173 | 5158 | 4603 | 7329 | 7481 | 3195 | 2191 | 5768 | 4655 | 0.62 | 0.39 | 1.57 | 0.66 | 0.52 | 1.29 | 87.00 | 70.23 | 55.95 |
| YML128C | 338 | 313 | 52 | 47 | 1327 | 1519 | 114 | 174 | 58 | 80 | 76 | 147 | 365 | 567 | 249 | 569 | 6.58 | 9.88 | 0.66 | 0.62 | 1.14 | 0.54 | 16.88 | 5.70 | 75.52 |
| YBR149W | 2925 | 2088 | 1412 | 1034 | 2895 | 3234 | 755 | 1019 | 1077 | 1571 | 992 | 1394 | 1366 | 2033 | 1071 | 1841 | 2.05 | 3.45 | 0.59 | 1.11 | 1.17 | 0.95 | 14.52 | 12.47 | 9.71 |
| YLR058C | 6655 | 5402 | 4563 | 7235 | 5393 | 6405 | 6182 | 10964 | 10387 | 29028 | 10235 | 26021 | 15605 | 34334 | 13911 | 26066 | 1.02 | 0.69 | 1.44 | 1.09 | 1.25 | 0.86 | 57.47 | 27.17 | 22.27 |
| YOL158C | 571 | 176 | 1539 | 1096 | 1960 | 1646 | 4060 | 4273 | 788 | 1188 | 641 | 498 | 2159 | 4747 | 2722 | 4742 | 0.28 | 0.43 | 0.63 | 1.73 | 0.93 | 1.96 | 23.32 | 9.28 | 37.11 |
| YPL012W | 3367 | 1588 | 5456 | 3199 | 1661 | 1335 | 3033 | 3765 | 2909 | 2860 | 2882 | 2969 | 2102 | 1032 | 1974 | 1609 | 0.57 | 0.44 | 1.23 | 0.99 | 0.87 | 1.15 | 2.53 | 16.34 | 67.86 |
| YGR086C | 9686 | 6874 | 5447 | 3866 | 7994 | 8310 | 3448 | 4433 | 5456 | 6788 | 5160 | 6491 | 6143 | 5993 | 5140 | 4722 | 1.78 | 2.07 | 0.85 | 1.05 | 1.23 | 0.85 | 8.63 | 28.54 | 97.81 |
| YNL061W | 2327 | 1292 | 4783 | 2807 | 1712 | 1377 | 3201 | 3817 | 2477 | 2290 | 2475 | 2460 | 2076 | 992 | 2300 | 1233 | 0.48 | 0.44 | 1.05 | 0.97 | 0.87 | 1.13 | 4.68 | 17.20 | 62.29 |
| YGL078C | 1935 | 1166 | 3903 | 2462 | 1689 | 1402 | 3920 | 3850 | 1942 | 1767 | 2277 | 2268 | 1690 | 727 | 2606 | 973 | 0.49 | 0.40 | 1.21 | 0.82 | 0.68 | 1.21 | 28.27 | 42.59 | 99.21 |
| YDL078C | 1835 | 1317 | 387 | 260 | 1798 | 1851 | 270 | 325 | 1098 | 1281 | 399 | 511 | 1377 | 1307 | 577 | 792 | 4.87 | 6.13 | 0.79 | 2.61 | 1.96 | 1.34 | 60.69 | 37.12 | 35.74 |
| YMR189W | 3249 | 2301 | 129 | 153 | 2765 | 3201 | 650 | 1026 | 568 | 2059 | 482 | 1917 | 1731 | 5623 | 1849 | 7040 | 19.68 | 3.56 | 5.28 | 1.10 | 0.83 | 1.33 | 3.05 | 11.50 | 16.96 |
| YJR104C | 9454 | 8622 | 6013 | 3682 | 5661 | 5804 | 3534 | 2706 | 8454 | 9711 | 6047 | 7086 | 5661 | 4625 | 4018 | 3207 | 1.86 | 1.84 | 1.03 | 1.38 | 1.42 | 0.97 | 52.07 | 58.07 | 33.46 |
| YJR048W | 1179 | 683 | 4869 | 3284 | 524 | 441 | 2757 | 2076 | 4774 | 5757 | 4194 | 5087 | 3644 | 2655 | 2739 | 2465 | 0.23 | 0.20 | 1.14 | 1.13 | 1.21 | 0.95 | 7.31 | 9.58 | 23.00 |
| YMR319C | 855 | 434 | 117 | 146 | 2669 | 1831 | 274 | 328 | 367 | 347 | 189 | 223 | 1233 | 2021 | 394 | 688 | 4.90 | 7.48 | 0.63 | 1.73 | 3.01 | 0.57 | 34.59 | 54.74 | 84.69 |
| YOR230W | 7963 | 6485 | 6698 | 4720 | 7987 | 9981 | 4264 | 5756 | 4380 | 6076 | 4729 | 6467 | 5898 | 10118 | 6007 | 9536 | 1.27 | 1.79 | 0.71 | 0.93 | 1.03 | 0.91 | 18.38 | 5.13 | 27.89 |
| YBL054W | 332 | 174 | 900 | 683 | 879 | 707 | 1631 | 2592 | 527 | 297 | 469 | 259 | 765 | 1108 | 1247 | 1211 | 0.32 | 0.38 | 0.77 | 1.13 | 0.76 | 1.57 | 8.92 | 27.75 | 38.87 |
| YBL076C | 17787 | 9684 | 15409 | 8692 | 5889 | 5533 | 7818 | 9637 | 11032 | 12368 | 11249 | 12004 | 7161 | 7035 | 7423 | 9402 | 1.14 | 0.65 | 1.71 | 1.01 | 0.84 | 1.19 | 4.82 | 40.06 | 32.86 |
| YPR074C | 22877 | 15062 | 25326 | 14338 | 6769 | 6528 | 9443 | 10930 | 15486 | 17802 | 14892 | 17144 | 8105 | 12615 | 8708 | 12134 | 0.96 | 0.65 | 1.47 | 1.04 | 0.99 | 1.05 | 31.65 | 1.38 | 12.04 |
| YMR300C | 1259 | 795 | 1629 | 2177 | 1383 | 1393 | 2223 | 3750 | 2280 | 5338 | 2036 | 5148 | 5194 | 3953 | 4097 | 4505 | 0.54 | 0.46 | 1.11 | 1.06 | 1.06 | 1.06 | 7.99 | 6.91 | 53.61 |
| YLR256W | 2253 | 1506 | 573 | 384 | 3134 | 2886 | 806 | 885 | 795 | 864 | 664 | 784 | 1463 | 1659 | 1101 | 1392 | 3.93 | 3.56 | 1.09 | 1.15 | 1.25 | 0.91 | 9.94 | 17.72 | 33.56 |
| YLR432W | 11782 | 7619 | 3901 | 2663 | 7204 | 6480 | 2769 | 3220 | 4493 | 5003 | 3489 | 4108 | 3830 | 4781 | 3414 | 4315 | 2.96 | 2.28 | 1.28 | 1.25 | 1.11 | 1.12 | 20.59 | 13.08 | 46.89 |
| YCR057C | 1709 | 1061 | 2849 | 1706 | 1374 | 1192 | 2246 | 3412 | 1858 | 1568 | 1678 | 1641 | 1780 | 2384 | 1831 | 2309 | 0.61 | 0.45 | 1.27 | 1.03 | 1.01 | 1.02 | 5.66 | 0.72 | 9.50 |
| YBR214W | 721 | 622 | 291 | 203 | 1413 | 1656 | 157 | 267 | 267 | 481 | 239 | 528 | 390 | 1356 | 357 | 963 | 2.72 | 7.24 | 0.37 | 0.98 | 1.32 | 0.74 | 2.39 | 14.13 | 29.84 |
| YNL141W | 1053 | 557 | 3656 | 2464 | 1337 | 1173 | 2914 | 3379 | 1249 | 1024 | 2170 | 2008 | 2223 | 1248 | 2563 | 2123 | 0.26 | 0.40 | 0.64 | 0.54 | 0.74 | 0.75 | 45.59 | 32.65 | 65.33 |

| | Adjusted P-values for tests of divergent expression | | | | | Adjusted P-values for *cis*-regulatory divergence in F1 Hybrids | | | | | Adjusted P-values for tests of *trans*-regulatory divergence | | | tAI | |
|---|---|---|---|---|---|---|---|---|---|---|---|---|---|---|---|
| | mRNA | | RPF | | Eff | mRNA | | RPF | | Eff | Eff | RPF | mRNA | | |
| Gene | Rep A | Rep B | Rep A | Rep B | Combined | Rep A | Rep B | Rep A | Rep B | Combined | Trans_eff_pad | RPF_trans_pa | mRNA_trans_ | *S. cer* | *S. par* |
| YKR046C | 4490535E-311 | 0.00E+00 | 0.00E+00 | 0.00E+00 | 2.10E-02 | 7.48E-120 | 2.17E-124 | 1.63E-90 | 6.79E-148 | 1.35E-01 | 3.40E-01 | 4.29E-104 | 3.81E-89 | 0.44 | 0.43 |
| YHR071W | 920623E-322 | 0.00E+00 | 4.22E-31 | 3.74E-34 | 6.34E-10 | 4.44E-132 | 1.71E-104 | 1.29E-07 | 6.43E-03 | 4.40E-04 | 7.84E-01 | 3.27E-04 | 1.71E-32 | 0.36 | 0.35 |
| YOR317W | 920623E-322 | 0.00E+00 | 754273E-322 | 0.00E+00 | 2.47E-12 | 9.35E-87 | 2.36E-81 | 1.39E-39 | 4.49E-23 | 9.80E-08 | 6.23E-02 | 4.72E-255 | 928024E-312 | 0.45 | 0.44 |
| YIL051C | 920623E-322 | 0.00E+00 | 754273E-322 | 3.45E-271 | 1.88E-37 | 1.24E-09 | 5.40E-03 | 1.50E-21 | 1.35E-47 | 7.38E-39 | 2.24E-01 | 0.00E+00 | 0.00E+00 | 0.48 | 0.47 |
| YDR481C | 920623E-322 | 0.00E+00 | 754273E-322 | 0.00E+00 | 6.88E-07 | 2.22E-07 | 2.52E-04 | 7.11E-01 | 3.33E-03 | 6.50E-07 | 8.84E-01 | 0.00E+00 | 0.00E+00 | 0.40 | 0.40 |
| YOR273C | 920623E-322 | 0.00E+00 | 1.97E-08 | 3.40E-30 | 2.67E-199 | 3.05E-272 | 0.00E+00 | 1.66E-48 | 2.96E-190 | 1.58E-33 | 1.30E-148 | 8.17E-37 | 6.13E-01 | 0.40 | 0.40 |
| YDR533C | 967688E-318 | 0.00E+00 | 9.96E-178 | 1.12E-223 | 9.62E-03 | 8.65E-43 | 1.18E-67 | 1.09E-23 | 1.00E-41 | 7.69E-01 | 1.82E-02 | 1.14E-14 | 3.26E-46 | 0.44 | 0.47 |
| YPL126W | 3.70E-67 | 0.00E+00 | 1.43E-13 | 2.48E-43 | 1.00E-50 | 1.51E-01 | 1.27E-01 | 8.13E-05 | 7.49E-04 | 6.34E-02 | 7.85E-10 | 1.02E-45 | 4.13E-150 | 0.39 | 0.39 |
| YMR058W | 2.48E-77 | 0.00E+00 | 5.62E-70 | 088873E-322 | 3.54E-11 | 5.97E-79 | 9.25E-189 | 2.40E-246 | 335359E-321 | 1.27E-05 | 5.73E-13 | 6.63E-01 | 1.10E-35 | 0.41 | 0.41 |
| YLR214W | 7.15E-94 | 0.00E+00 | 1.60E-94 | 904755E-316 | 9.87E-01 | 6.99E-14 | 2.32E-98 | 8.59E-41 | 3.42E-35 | 9.56E-03 | 7.06E-02 | 2.33E-05 | 2.59E-03 | 0.35 | 0.35 |
| YFR055W | 5.45E-130 | 0.00E+00 | 754273E-322 | 0.00E+00 | 4.83E-07 | 1.20E-36 | 1.95E-103 | 1.96E-34 | 5.49E-33 | 5.60E-05 | 8.09E-11 | 7.13E-163 | 2.37E-92 | 0.36 | 0.34 |
| YMR290C | 4.14E-137 | 0.00E+00 | 3.37E-138 | 4.97E-133 | 1.15E-17 | 1.44E-60 | 1.86E-06 | 2.01E-07 | 2.87E-09 | 5.84E-10 | 3.88E-01 | 4.51E-76 | 1.10E-71 | 0.45 | 0.42 |
| YMR305C | 3.09E-164 | 0.00E+00 | 7.31E-137 | 3.24E-91 | 8.32E-21 | 1.10E-13 | 9.86E-27 | 1.11E-11 | 7.38E-10 | 1.44E-01 | 2.35E-05 | 1.53E-39 | 3.13E-115 | 0.45 | 0.46 |
| YJL198W | 4.77E-174 | 0.00E+00 | 4.21E-39 | 5.02E-30 | 1.54E-26 | 6.35E-42 | 3.22E-21 | 4.77E-02 | 3.55E-04 | 4.72E-06 | 3.70E-03 | 1.44E-18 | 2.10E-117 | 0.40 | 0.38 |
| YBR084W | 1.63E-190 | 0.00E+00 | 1.16E-55 | 4.36E-70 | 3.48E-88 | 5.53E-20 | 1.98E-223 | 6.41E-01 | 6.36E-02 | 3.80E-75 | 2.17E-01 | 1.37E-51 | 1.50E-68 | 0.43 | 0.42 |
| YJL196C | 8.04E-194 | 0.00E+00 | 1.10E-13 | 4.43E-34 | 3.74E-146 | 8.13E-85 | 2.20E-132 | 3.29E-02 | 4.39E-03 | 1.98E-77 | 2.18E-06 | 1.31E-07 | 2.26E-14 | 0.37 | 0.35 |
| YGR175C | 1.67E-202 | 0.00E+00 | 1.29E-287 | 1.15E-124 | 1.27E-45 | 9.35E-01 | 9.70E-01 | 1.15E-06 | 4.02E-04 | 9.98E-05 | 9.46E-10 | 2.53E-266 | 0.00E+00 | 0.52 | 0.49 |
| YOR374W | 1.27E-203 | 0.00E+00 | 0.00E+00 | 3.59E-266 | 7.56E-01 | 8.90E-10 | 8.86E-12 | 5.25E-45 | 2.78E-36 | 4.11E-09 | 6.24E-04 | 5.14E-152 | 1.07E-199 | 0.47 | 0.47 |
| YBR067C | 8.84E-221 | 0.00E+00 | 3.41E-303 | 0.00E+00 | 8.91E-02 | 7.16E-02 | 4.54E-02 | 7.41E-01 | 6.55E-01 | 9.91E-01 | 3.38E-01 | 1.15E-203 | 4.05E-187 | 0.53 | 0.57 |
| YGL037C | 3.15E-225 | 0.00E+00 | 8.35E-295 | 9.14E-147 | 1.11E-08 | 2.68E-01 | 1.06E-12 | 8.60E-01 | 4.79E-03 | 3.91E-03 | 6.70E-02 | 3.76E-156 | 1.52E-173 | 0.47 | 0.47 |
| YGR088W | 8.26E-239 | 0.00E+00 | 7.87E-03 | 9.77E-11 | 8.37E-32 | 1.18E-03 | 4.14E-04 | 6.49E-04 | 9.85E-02 | 1.23E-06 | 1.37E-33 | 1.29E-14 | 8.95E-227 | 0.43 | 0.44 |
| YMR006C | 4.86E-252 | 0.00E+00 | 0.00E+00 | 0.00E+00 | 4.02E-02 | 6.97E-90 | 710597E-321 | 1.64E-24 | 1.41E-118 | 5.70E-27 | 2.90E-05 | 1.63E-139 | 1.19E-49 | 0.41 | 0.43 |
| YDL039C | 5.56E-276 | 0.00E+00 | 1.95E-137 | 1.63E-100 | 1.11E-13 | 2.41E-09 | 2.23E-42 | 5.35E-28 | 5.03E-02 | 1.21E-40 | 4.74E-02 | 2.02E-36 | 0.00E+00 | 0.38 | 0.40 |
| YNL111C | 3.18E-290 | 0.00E+00 | 0.00E+00 | 0.00E+00 | 1.09E-02 | 6.33E-01 | 1.00E+00 | 8.34E-01 | 7.91E-01 | 6.35E-01 | 3.30E-02 | 0.00E+00 | 0.00E+00 | 0.49 | 0.49 |
| YMR145C | 8.44E-298 | 0.00E+00 | 0.00E+00 | 3.19E-209 | 6.59E-05 | 2.46E-01 | 2.40E-01 | 8.06E-01 | 4.54E-01 | 6.71E-01 | 2.12E-03 | 2.33E-305 | 0.00E+00 | 0.45 | 0.43 |
| YHR143W | 5.62E-301 | 0.00E+00 | 0.00E+00 | 2.84E-221 | 2.72E-06 | 2.19E-50 | 3.13E-94 | 1.64E-06 | 8.71E-02 | 9.69E-46 | 4.87E-17 | 2.47E-192 | 1.76E-83 | 0.39 | 0.40 |
| YML075C | 0.00E+00 | 0.00E+00 | 0.00E+00 | 088873E-322 | 3.82E-05 | 1.61E-58 | 3.50E-110 | 5.00E-105 | 1.30E-89 | 2.58E-04 | 2.54E-01 | 0.00E+00 | 0.00E+00 | 0.41 | 0.41 |
| YBR092C | 0.00E+00 | 0.00E+00 | 0.00E+00 | 0.00E+00 | 6.21E-24 | 7.68E-01 | 1.93E-94 | 3.49E-138 | 1.32E-218 | 0.00E+00 | 1.89E-02 | 3.57E-115 | 0.00E+00 | 0.49 | 0.45 |
| YHR136C | 0.00E+00 | 0.00E+00 | 0.00E+00 | 3.07E-304 | 2.87E-03 | 2.99E-01 | 5.37E-01 | 8.65E-01 | 4.81E-02 | 2.79E-01 | 1.65E-01 | 0.00E+00 | 0.00E+00 | 0.35 | 0.40 |
| YML100W | 0.00E+00 | 0.00E+00 | 0.00E+00 | 0.00E+00 | 9.89E-02 | 4.39E-04 | 1.50E-10 | 1.98E-03 | 7.54E-03 | 2.51E-06 | 1.22E-04 | 5.27E-294 | 0.00E+00 | 0.39 | 0.39 |
| YMR121C | 0.00E+00 | 0.00E+00 | 0.00E+00 | 088873E-322 | 3.62E-07 | 0.00E+00 | 710597E-321 | 0.00E+00 | 335359E-321 | 5.23E-11 | 2.71E-01 | 1.04E-11 | 8.77E-19 | 0.55 | 0.56 |
| YNL036W | 0.00E+00 | 0.00E+00 | 754273E-322 | 6.91E-226 | 3.61E-20 | 3.91E-18 | 1.52E-30 | 2.32E-01 | 3.91E-07 | 1.47E-06 | 6.33E-09 | 0.00E+00 | 0.00E+00 | 0.41 | 0.41 |
| YIL094C | 0.00E+00 | 0.00E+00 | 0.00E+00 | 667625E-316 | 4.95E-06 | 2.47E-285 | 2.73E-162 | 0.00E+00 | 798971E-318 | 7.45E-06 | 3.53E-01 | 7.93E-54 | 7.66E-53 | 0.49 | 0.47 |
| YDR001C | 0.00E+00 | 0.00E+00 | 0.00E+00 | 2.57E-249 | 5.52E-01 | 7.70E-163 | 1.99E-128 | 9.93E-44 | 6.58E-92 | 4.61E-02 | 5.55E-01 | 1.28E-57 | 4.61E-75 | 0.39 | 0.37 |
| YDL192W | 0.00E+00 | 0.00E+00 | 0.00E+00 | 0.00E+00 | 8.51E-06 | 2.73E-295 | 9.52E-55 | 0.00E+00 | 0.00E+00 | 1.10E-23 | 2.11E-03 | 4.00E-02 | 9.50E-14 | 0.53 | 0.50 |
| YDL124W | 0.00E+00 | 0.00E+00 | 0.00E+00 | 5.47E-264 | 4.74E-01 | 2.27E-02 | 2.10E-05 | 6.72E-09 | 2.26E-16 | 1.61E-07 | 2.29E-04 | 1.19E-95 | 7.26E-279 | 0.43 | 0.40 |
| YER072W | 0.00E+00 | 0.00E+00 | 754273E-322 | 0.00E+00 | 1.11E-08 | 1.24E-05 | 3.93E-66 | 1.08E-06 | 1.42E-05 | 1.70E-09 | 9.58E-15 | 0.00E+00 | 0.00E+00 | 0.36 | 0.34 |
| YNR060W | 0.00E+00 | 0.00E+00 | 0.00E+00 | 0.00E+00 | 5.84E-16 | 9.12E-07 | 3.50E-01 | 5.77E-05 | 7.67E-01 | 9.73E-01 | 1.49E-46 | 0.00E+00 | 0.00E+00 | 0.38 | 0.37 |
| YDR074W | 0.00E+00 | 0.00E+00 | 7.23E-87 | 1.42E-95 | 7.09E-173 | 1.20E-23 | 3.30E-08 | 5.80E-07 | 7.77E-07 | 8.45E-01 | 1.80E-204 | 4.01E-142 | 0.00E+00 | 0.42 | 0.42 |
| YKL035W | 0.00E+00 | 0.00E+00 | 8.99E-149 | 4.71E-152 | 6.66E-78 | 3.13E-07 | 1.85E-31 | 6.32E-01 | 3.04E-02 | 5.86E-10 | 2.66E-32 | 5.01E-122 | 864679E-320 | 0.49 | 0.46 |
| YMR195W | 0.00E+00 | 0.00E+00 | 0.00E+00 | 3.84E-210 | 1.23E-06 | 7.43E-02 | 5.52E-01 | 7.33E-01 | 4.04E-02 | 4.06E-01 | 1.45E-04 | 1.60E-288 | 0.00E+00 | 0.36 | 0.40 |
| YMR250W | 0.00E+00 | 0.00E+00 | 4.19E-124 | 9.98E-84 | 7.28E-14 | 3.29E-02 | 8.43E-11 | 8.89E-01 | 9.32E-02 | 1.80E-04 | 9.71E-09 | 1.90E-81 | 0.00E+00 | 0.40 | 0.38 |
| YMR303C | 0.00E+00 | 0.00E+00 | 0.00E+00 | 0.00E+00 | 5.18E-03 | 878846E-321 | 0.00E+00 | 252564E-321 | 335359E-321 | 1.95E-19 | 1.18E-01 | 4.69E-01 | 2.48E-06 | 0.56 | 0.52 |
| YGR213C | 0.00E+00 | 0.00E+00 | 2.05E-177 | 3.60E-191 | 1.52E-05 | 0.00E+00 | 0.00E+00 | 5.25E-95 | 3.50E-114 | 4.10E-03 | 3.05E-01 | 3.27E-01 | 2.80E-01 | 0.35 | 0.37 |
| YOR161C | 0.00E+00 | 0.00E+00 | 2.61E-124 | 5.66E-126 | 6.84E-16 | 4.48E-47 | 1.31E-83 | 1.94E-13 | 2.36E-04 | 3.12E-07 | 2.57E-10 | 3.19E-56 | 4.16E-189 | 0.37 | 0.37 |
| YAR015W | 0.00E+00 | 0.00E+00 | 1.01E-75 | 088873E-322 | 1.29E-25 | 1.62E-98 | 7.04E-45 | 2.14E-38 | 1.73E-40 | 1.42E-01 | 2.15E-08 | 1.27E-157 | 0.00E+00 | 0.45 | 0.44 |
| YDL022W | 0.00E+00 | 0.00E+00 | 2.26E-149 | 3.26E-215 | 2.59E-124 | 1.39E-32 | 8.55E-59 | 7.66E-75 | 4.56E-83 | 1.08E-01 | 9.12E-122 | 1.07E-13 | 4.92E-193 | 0.53 | 0.53 |
| YLR258W | 0.00E+00 | 0.00E+00 | 667486E-319 | 3.18E-212 | 3.04E-01 | 8.11E-05 | 1.23E-03 | 3.02E-01 | 1.01E-02 | 7.92E-01 | 3.58E-01 | 6.35E-214 | 0.00E+00 | 0.41 | 0.39 |
| YHL021C | 0.00E+00 | 0.00E+00 | 3.98E-194 | 3.93E-169 | 5.55E-51 | 2.47E-03 | 3.69E-10 | 7.84E-03 | 3.07E-08 | 2.38E-01 | 8.16E-78 | 5.17E-128 | 0.00E+00 | 0.38 | 0.38 |
| YJL052W | 920623E-322 | 1.10E-307 | 754273E-322 | 8.64E-209 | 1.04E-14 | 878846E-321 | 710597E-321 | 252564E-321 | 335359E-321 | 3.42E-06 | 6.89E-01 | 0.00E+00 | 0.00E+00 | 0.64 | 0.62 |
| YJR148W | 0.00E+00 | 2.64E-307 | 1.01E-148 | 2.24E-97 | 2.09E-60 | 1.15E-273 | 710597E-321 | 252564E-321 | 335359E-321 | 6.49E-02 | 1.64E-42 | 9.18E-86 | 6.73E-02 | 0.44 | 0.44 |
| YNL160W | 2.83E-276 | 1.55E-305 | 8.27E-302 | 1.57E-298 | 4.69E-21 | 1.21E-17 | 1.41E-49 | 2.16E-01 | 6.24E-02 | 1.34E-21 | 5.47E-01 | 1.38E-251 | 0.00E+00 | 0.51 | 0.52 |
| YLR355C | 9.60E-160 | 1.37E-303 | 754273E-322 | 088873E-322 | 2.40E-02 | 3.26E-04 | 1.82E-01 | 1.84E-44 | 3.54E-43 | 3.02E-12 | 7.12E-10 | 0.00E+00 | 8.77E-271 | 0.62 | 0.58 |
| YBR052C | 1.92E-255 | 6.56E-303 | 2.68E-224 | 1.53E-204 | 4.65E-04 | 5.74E-85 | 1.97E-85 | 6.70E-41 | 1.23E-60 | 2.08E-05 | 9.24E-01 | 6.56E-47 | 8.53E-41 | 0.41 | 0.41 |
| YGR155W | 9.67E-90 | 1.04E-299 | 3.81E-02 | 5.40E-06 | 1.90E-217 | 2.19E-42 | 2.05E-17 | 1.31E-01 | 1.24E-05 | 9.89E-50 | 2.08E-16 | 1.07E-01 | 2.59E-67 | 0.50 | 0.49 |
| YMR307W | 1.02E-296 | 1.23E-298 | 3.63E-08 | 7.37E-06 | 0.00E+00 | 8.02E-05 | 8.19E-13 | 1.63E-02 | 1.09E-04 | 2.00E-03 | 3.24E-99 | 1.71E-16 | 2.76E-217 | 0.49 | 0.50 |
| YLR409C | 1.06E-103 | 5.77E-296 | 1.55E-79 | 8.83E-47 | 8.84E-09 | 2.63E-16 | 2.32E-120 | 2.24E-01 | 2.53E-02 | 1.53E-28 | 1.54E-02 | 1.24E-47 | 1.27E-37 | 0.40 | 0.41 |
| YDR232W | 4.59E-224 | 1.58E-292 | 2.03E-118 | 5.91E-35 | 8.08E-122 | 1.11E-47 | 4.38E-25 | 3.69E-07 | 2.49E-01 | 1.89E-26 | 5.24E-07 | 4.57E-48 | 1.58E-135 | 0.43 | 0.43 |
| YMR011W | 3.49E-290 | 1.01E-290 | 754273E-322 | 088873E-322 | 1.37E-46 | 6.37E-172 | 2.98E-235 | 252564E-321 | 335359E-321 | 9.38E-16 | 4.10E-26 | 0.00E+00 | 0.00E+00 | 0.48 | 0.49 |
| YDR055W | 920623E-322 | 1.29E-286 | 2.23E-46 | 1.48E-08 | 1.31E-124 | 2.45E-89 | 6.55E-70 | 3.60E-09 | 1.75E-04 | 3.66E-15 | 1.75E-44 | 2.87E-03 | 4.11E-94 | 0.44 | 0.43 |
| YBR177C | 4.17E-286 | 2.06E-286 | 4.24E-169 | 2.38E-114 | 2.55E-09 | 4.04E-04 | 9.59E-03 | 3.60E-07 | 1.97E-01 | 6.00E-01 | 3.28E-07 | 1.52E-185 | 0.00E+00 | 0.39 | 0.40 |
| YJL079C | 3.57E-211 | 5.49E-286 | 9.32E-252 | 670718E-321 | 5.60E-01 | 2.27E-03 | 4.86E-24 | 7.22E-04 | 3.32E-10 | 4.97E-01 | 4.00E-01 | 1.56E-136 | 6.86E-117 | 0.47 | 0.44 |
| YNL190W | 2.15E-157 | 9.49E-281 | 5.03E-18 | 1.06E-24 | 1.54E-158 | 1.45E-217 | 6.19E-160 | 4.97E-04 | 1.51E-01 | 4.27E-185 | 9.31E-01 | 6.79E-11 | 1.86E-08 | 0.56 | 0.57 |
| YBL003C | 4.28E-125 | 1.88E-280 | 2.80E-247 | 2.00E-237 | 6.78E-23 | 1.45E-12 | 9.09E-20 | 2.32E-20 | 1.46E-17 | 4.02E-01 | 2.13E-07 | 1.65E-134 | 6.66E-72 | 0.56 | 0.53 |
| YDR225W | 1.24E-287 | 1.98E-280 | 2.63E-106 | 5.78E-177 | 3.83E-06 | 2.91E-02 | 1.11E-09 | 5.01E-01 | 9.72E-01 | 8.49E-04 | 8.09E-09 | 5.28E-154 | 0.00E+00 | 0.60 | 0.55 |
| YBR026C | 2.18E-254 | 3.80E-279 | 4.23E-190 | 1.97E-129 | 1.90E-04 | 2.80E-36 | 1.67E-33 | 7.78E-37 | 1.88E-26 | 7.92E-01 | 4.37E-03 | 1.57E-37 | 1.27E-68 | 0.39 | 0.40 |
| YOR264W | 0.00E+00 | 1.32E-278 | 1.32E-65 | 5.13E-32 | 3.95E-13 | 5.19E-30 | 2.42E-12 | 1.36E-07 | 2.58E-29 | 5.21E-02 | 6.26E-11 | 8.23E-06 | 2.36E-85 | 0.37 | 0.35 |
| YOR128C | 6.74E-40 | 7.73E-278 | 7.60E-08 | 6.47E-106 | 3.11E-09 | 1.56E-07 | 5.22E-34 | 2.83E-10 | 4.72E-12 | 9.32E-58 | 7.99E-36 | 7.57E-33 | 4.88E-285 | 0.41 | 0.41 |
| YMR205C | 6.60E-191 | 2.68E-276 | 3.95E-25 | 5.51E-41 | 0.00E+00 | 4.49E-08 | 2.07E-11 | 4.59E-01 | 8.70E-06 | 1.70E-06 | 1.79E-61 | 7.18E-47 | 6.97E-197 | 0.55 | 0.54 |
| YLR134W | 6.75E-188 | 5.49E-276 | 6.87E-82 | 5.48E-157 | 1.40E-60 | 4.61E-187 | 3.11E-297 | 7.95E-80 | 8.35E-47 | 1.71E-100 | 2.20E-01 | 6.37E-11 | 2.59E-02 | 0.60 | 0.56 |
| YGL009C | 8.64E-55 | 3.09E-275 | 6.59E-44 | 2.13E-123 | 3.70E-35 | 1.58E-55 | 1.19E-67 | 1.49E-04 | 6.23E-32 | 6.76E-30 | 5.16E-01 | 7.52E-27 | 6.06E-23 | 0.49 | 0.49 |
| YHR006W | 3.87E-236 | 3.57E-274 | 1.11E-12 | 5.23E-05 | 4.30E-67 | 5.10E-22 | 7.14E-32 | 8.47E-01 | 3.22E-04 | 1.45E-27 | 4.72E-18 | 2.10E-12 | 5.46E-67 | 0.34 | 0.34 |
| YGR060W | 6.63E-209 | 1.89E-272 | 3.58E-191 | 3.13E-175 | 1.51E-21 | 1.50E-01 | 4.16E-13 | 2.11E-01 | 2.96E-01 | 7.78E-03 | 4.86E-05 | 8.05E-179 | 2.87E-197 | 0.50 | 0.48 |
| YDL131W | 1.30E-287 | 9.33E-272 | 3.50E-284 | 4.37E-156 | 1.11E-01 | 1.80E-30 | 1.76E-19 | 4.31E-15 | 1.10E-23 | 9.99E-01 | 2.21E-01 | 1.06E-180 | 8.92E-218 | 0.47 | 0.46 |
| YDL063C | 7.77E-85 | 1.56E-271 | 2.30E-134 | 4.66E-148 | 1.56E-11 | 2.63E-15 | 6.24E-24 | 2.61E-07 | 3.83E-10 | 1.04E-01 | 3.29E-10 | 6.32E-73 | 1.90E-37 | 0.35 | 0.36 |
| YMR229C | 920623E-322 | 2.89E-270 | 1.31E-72 | 2.89E-76 | 2.62E-224 | 1.46E-15 | 1.43E-02 | 1.38E-01 | 3.17E-01 | 1.52E-08 | 7.79E-23 | 9.50E-62 | 6.23E-191 | 0.46 | 0.46 |
| YJL109C | 1.65E-218 | 3.78E-270 | 3.23E-87 | 2.03E-104 | 5.83E-56 | 3.20E-01 | 8.23E-20 | 6.47E-05 | 2.35E-08 | 5.61E-01 | 1.92E-15 | 1.44E-49 | 4.79E-166 | 0.40 | 0.39 |
| YDR502C | 8.36E-192 | 7.72E-270 | 1.08E-114 | 1.25E-101 | 1.64E-62 | 1.14E-162 | 4.48E-197 | 1.27E-82 | 1.93E-150 | 8.31E-31 | 2.50E-04 | 9.25E-04 | 1.63E-21 | 0.54 | 0.53 |
| YML128C | 5.68E-261 | 1.57E-266 | 1.29E-51 | 2.04E-48 | 1.62E-03 | 9.15E-06 | 9.88E-01 | 2.15E-01 | 3.71E-05 | 9.63E-07 | 8.99E-02 | 6.17E-59 | 3.27E-189 | 0.44 | 0.44 |
| YBR149W | 3.30E-291 | 7.22E-264 | 6.71E-118 | 1.21E-79 | 1.01E-46 | 9.62E-09 | 3.91E-03 | 1.11E-01 | 3.01E-03 | 2.55E-01 | 3.53E-29 | 6.67E-58 | 3.22E-189 | 0.44 | 0.43 |
| YLR058C | 4.86E-13 | 1.16E-263 | 3.25E-86 | 6.55E-59 | 6.33E-92 | 6.28E-22 | 1.13E-246 | 3.92E-01 | 3.08E-36 | 1.66E-47 | 9.69E-78 | 4.48E-08 | 0.00E+00 | 0.56 | 0.54 |
| YOL158C | 2.82E-163 | 7.10E-263 | 5.81E-101 | 3.12E-161 | 9.46E-23 | 5.05E-15 | 9.82E-01 | 3.75E-04 | 1.94E-63 | 1.10E-58 | 2.31E-38 | 2.49E-240 | 1.91E-181 | 0.33 | 0.32 |
| YPL012W | 2.45E-89 | 1.49E-262 | 2.39E-109 | 8.47E-121 | 1.03E-12 | 6.99E-02 | 2.16E-28 | 8.07E-01 | 2.26E-01 | 1.85E-05 | 2.52E-01 | 5.01E-103 | 6.15E-81 | 0.41 | 0.40 |
| YGR086C | 920623E-322 | 5.65E-261 | 5.14E-262 | 4.76E-186 | 6.95E-17 | 3.26E-20 | 1.28E-33 | 9.91E-03 | 2.03E-02 | 3.14E-16 | 9.09E-01 | 9.11E-177 | 8.58E-161 | 0.52 | 0.50 |
| YNL061W | 1.75E-100 | 1.28E-259 | 8.46E-189 | 3.11E-125 | 1.48E-01 | 1.55E-03 | 9.97E-07 | 1.00E+00 | 2.73E-02 | 7.19E-04 | 1.93E-01 | 7.49E-135 | 3.02E-80 | 0.47 | 0.48 |
| YGL078C | 7.26E-199 | 2.83E-258 | 1.02E-147 | 1.51E-103 | 1.60E-09 | 2.97E-43 | 9.09E-09 | 1.36E-06 | 3.18E-14 | 3.81E-07 | 9.85E-01 | 8.56E-62 | 3.90E-50 | 0.48 | 0.49 |
| YDL078C | 5.81E-275 | 4.20E-257 | 3.77E-223 | 1.45E-168 | 3.53E-04 | 1.71E-73 | 1.68E-28 | 7.39E-74 | 1.82E-74 | 6.08E-08 | 2.92E-09 | 1.62E-25 | 5.40E-98 | 0.39 | 0.40 |
| YMR189W | 1.29E-306 | 1.08E-255 | 0.00E+00 | 088873E-322 | 3.11E-158 | 7.52E-02 | 2.69E-35 | 1.90E-02 | 4.59E-02 | 4.14E-17 | 3.68E-04 | 0.00E+00 | 0.00E+00 | 0.40 | 0.39 |
| YJR104C | 9.06E-109 | 2.53E-251 | 1.85E-168 | 088873E-322 | 2.09E-01 | 2.86E-61 | 1.15E-56 | 2.20E-87 | 9.00E-90 | 2.13E-01 | 7.38E-02 | 2.06E-74 | 2.85E-29 | 0.51 | 0.52 |
| YJR048W | 0.00E+00 | 5.91E-251 | 0.00E+00 | 0.00E+00 | 4.50E-03 | 1.18E-28 | 1.37E-02 | 6.73E-09 | 8.94E-10 | 3.62E-02 | 1.54E-03 | 0.00E+00 | 0.00E+00 | 0.51 | 0.48 |
| YMR319C | 0.00E+00 | 6.01E-251 | 6.41E-138 | 2.47E-33 | 4.13E-08 | 1.04E-98 | 5.76E-149 | 3.76E-13 | 1.06E-06 | 3.43E-14 | 1.80E-01 | 4.68E-26 | 3.82E-52 | 0.40 | 0.41 |
| YOR230W | 2.46E-250 | 4.04E-250 | 6.72E-25 | 1.37E-61 | 7.64E-82 | 3.89E-01 | 7.81E-05 | 8.34E-04 | 1.32E-03 | 3.26E-07 | 1.75E-32 | 3.90E-63 | 7.77E-238 | 0.45 | 0.46 |
| YBL054W | 8.53E-51 | 3.17E-249 | 5.65E-60 | 6.85E-71 | 3.81E-05 | 5.37E-26 | 5.09E-02 | 1.19E-01 | 1.75E-01 | 8.49E-13 | 1.70E-10 | 3.73E-69 | 5.86E-61 | 0.37 | 0.38 |
| YBL076C | 2.67E-60 | 6.97E-245 | 3.62E-38 | 9.08E-13 | 5.87E-274 | 4.76E-02 | 8.65E-75 | 2.23E-01 | 3.71E-02 | 1.86E-31 | 3.66E-29 | 2.04E-21 | 7.68E-53 | 0.50 | 0.49 |
| YPR074C | 3.23E-97 | 2.04E-244 | 3.33E-28 | 5.13E-05 | 7.35E-185 | 9.61E-06 | 4.13E-03 | 1.94E-03 | 1.18E-03 | 6.84E-04 | 2.55E-45 | 2.12E-10 | 5.76E-176 | 0.54 | 0.52 |
| YMR300C | 4.70E-44 | 3.23E-244 | 1.68E-11 | 1.11E-145 | 8.25E-03 | 6.74E-29 | 7.12E-09 | 6.87E-04 | 1.05E-01 | 3.50E-02 | 4.02E-01 | 3.47E-112 | 1.95E-181 | 0.44 | 0.44 |
| YLR256W | 147083E-319 | 1.90E-243 | 1.94E-232 | 3.95E-155 | 7.69E-02 | 4.74E-12 | 3.90E-06 | 1.92E-03 | 8.58E-02 | 8.37E-02 | 1.24E-02 | 1.77E-129 | 2.20E-163 | 0.37 | 0.38 |
| YLR432W | 920623E-322 | 2.42E-243 | 754273E-322 | 0.00E+00 | 1.78E-29 | 3.22E-06 | 2.94E-06 | 5.27E-28 | 9.13E-20 | 1.15E-06 | 1.14E-03 | 0.00E+00 | 3.96E-228 | 0.53 | 0.51 |
| YCR057C | 3.19E-47 | 4.51E-243 | 1.38E-63 | 6.02E-34 | 6.13E-12 | 4.72E-01 | 3.41E-01 | 6.53E-03 | 2.80E-01 | 5.93E-01 | 2.44E-04 | 1.57E-53 | 8.62E-135 | 0.39 | 0.40 |
| YBR214W | 2.55E-251 | 6.57E-243 | 6.28E-42 | 3.11E-49 | 1.24E-40 | 3.01E-01 | 1.84E-15 | 3.21E-01 | 2.14E-01 | 1.35E-05 | 1.21E-35 | 3.28E-42 | 2.21E-166 | 0.37 | 0.37 |
| YNL141W | 5.84E-131 | 7.11E-243 | 0.00E+00 | 1.29E-282 | 4.09E-33 | 2.85E-06 | 1.69E-50 | 7.45E-55 | 6.38E-71 | 1.99E-16 | 3.05E-05 | 6.22E-83 | 1.59E-71 | 0.44 | 0.45 |